\documentclass[preprint,showpacs,titlepage,aps,prd,tightenlines,amsmath,byrevtex,nofootinbib]{revtex4}

\usepackage{graphicx}
\usepackage{color}

\def\lsim{\raise0.3ex\hbox{$\;<$\kern-0.75em\raise-1.1ex
\hbox{$\sim\;$}}}
\def\gsim{\raise0.3ex\hbox{$\;>$\kern-0.75em\raise-1.1ex
\hbox{$\sim\;$}}}

\let\vev\VEV
\def\be{\begin{equation}}
\def\ee{\end{equation}}
\def\ba{\begin{eqnarray}}
 \def\ea{\end{eqnarray}}

\begin{document}

\preprint{IFIC/08-09}

\title{Parameter Degeneracy in Flavor-Dependent Reconstruction of
Supernova Neutrino Fluxes }

\vskip 2cm

\author{Hisakazu Minakata$^{1}$}
\email{minakata@phys.metro-u.ac.jp}
\author{Hiroshi Nunokawa$^{2}$}
\email{nunokawa@fis.puc-rio.br} 
\author{Ricard Tom{\`a}s$^{3}$\footnote{Present address: II. Institut
  f\"ur Theoretische Physik, Universit\"at Hamburg, 
Luruper Chaussee 149, 22761 Hamburg, Germany}} 
\email{ricard.tomas.bayo@desy.de} 
\author{Jose~W.~F. Valle$^{3}$}
\email{valle@ific.uv.es}
\affiliation{
$^1$Department of Physics, Tokyo Metropolitan University,  
Hachioji, Tokyo 192-0397, Japan \\
$^2$Departamento de F\'{\i}sica, Pontif{\'\i}cia Universidade Cat{\'o}lica 
do Rio de Janeiro, C. P. 38071, 22452-970, Rio de Janeiro, Brazil \\
$^3$AHEP Group, Institut de F\'{\i}sica Corpuscular -- C.S.I.C., 
Universitat de Val{\`e}ncia, Edifici Instituts d'Investigaci\'o, 
Apt.\ 22085, E--46071 Val{\`e}ncia, Spain 
}

\date{February 15, 2008}

\vglue 1.4cm

\begin{abstract}

  We reexamine the possibility of reconstructing the initial fluxes of
  supernova neutrinos emitted in a future core-collapse galactic
  supernova explosion and detected in a Megaton-sized water Cherenkov
  detector.  A novel   key element in our method is the inclusion,
  in addition to the total and the average energies of each neutrino
  species, of a ``pinching'' parameter characterizing the width of the
  distribution as a fit parameter.  We uncover in this case a
  continuous degeneracy in the reconstructed parameters of supernova neutrino
  fluxes at the neutrinosphere. 
  We analyze in detail the features of this
  degeneracy and show how it occurs irrespective of the
  parametrization used for the distribution function. Given that
  this degeneracy is real we briefly comment on possible steps towards
  resolving it, which necessarily requires going beyond the setting
  presented here.

\end{abstract}


\maketitle

\section{Introduction}

The detection of neutrinos from the SN1987A by the Kamiokande
\cite{Hirata:1987hu,Hirata:1988ad} and IMB \cite{Bionta:1987qt} 
collaborations not
only confirmed the basic picture of the core collapse supernovae (SN)
but also initiated a whole new field of observational supernova
astrophysics.
In fact, the observation of SN neutrinos is the unique way, 
except possibly for gravitational wave detection, to directly probe 
into the interior of a star which is undergoing gravitational collapse. 
Despite this pioneering observation, a precise understanding of the
physics of a SN explosion still eludes us.  Spherically symmetric
models of iron core collapse do not explode, even with solid neutrino
transport~\cite{Rampp:2000ws,Mezzacappa:2000jb,Thompson:2002mw} and
general relativity~\cite{Liebendoerfer:2000cq}. In the two-dimensional
models the outcomes vary qualitatively and
quantitatively~\cite{Buras:2005tb,Bruenn:2007bd,Burrows:2005dv,Cardall:2007ha},
reflecting the increased complexity of the physics involved.


Owing to their feeble interactions, neutrinos are able to escape 
from deep inside the star. Therefore SN neutrinos can provide us
with information about the highly dense inner layers, where the SN
explosion is initiated.
Moreover, the composition of the SN core is such that the reactions
involved in the creation and annihilation of neutrinos are different
for the different neutrino flavors. These differences are imprinted in
the emerging neutrino fluxes.  These are non-thermal and can be
characterized by the total energy emitted, the mean energy and the
so-called {\em pinching} parameter which controls the width of the
distribution. Therefore, a flavor-dependent reconstruction of SN
neutrino flux is a useful probe for diagnosing SN core.  (For early
references, see e.g., \cite{diagnostics}.)  The physics which can be
probed would include that governing matter under extreme conditions,
such as information about the equation of state, and the explosion
mechanism itself \cite{Raffelt:2007nv}.

The potential of neutrinos in probing the SN core results, on the one
hand, from the huge flux of them that will be released, corresponding
to around 99~\% of the total energy emitted in the SN explosion.
On the other hand it is helped (though it may sound ironic) by the
flavor mixing properties of neutrinos, including the large ``solar''
neutrino mixing angle obtained by KamLAND~\cite{Araki:2004mb} and the
solar neutrino observation, in particular, SNO~\cite{Aharmim:2005gt}.
For an updated global analysis of the data of the various neutrino
oscillation experiments, see e.g.,~\cite{Maltoni:2004ei,Schwetz:2008er}.

If a supernova explosion takes place in our galaxy the number of
neutrino events expected in the current and planned neutrino detectors
will be enormous~\cite{Burrows:1991kf,Scholberg:2007nu}.
Among all neutrino detectors water Cherenkov detectors are most likely
the ones which can run long enough to watch galactic SN over long
enough time scales.  It is therefore important to establish a strategy
for diagnosing the core of SN by using neutrino observation by water
Cherenkov detectors.

Inverse beta decay $\bar\nu_e + p \to e^+ + n$ provides the main
neutrino detection reaction in such a detector.
According to the standard picture of supernova neutrino propagation,
except for the case of inverted mass hierarchy and large $\theta_{13}$
($\sin^2\theta_{13} \gtrsim 10^{-3}$), the energy spectrum of
$\bar{\nu}_e$ at the Earth is expected to be a strong mixture of the
original $\bar{\nu}_e$ and $\nu_x$ (we collectively denote $\nu_\mu$,
$\nu_\tau$, $\bar{\nu}_\mu$, $\bar{\nu}_\tau$ as $\nu_x$ since at
first approximation their properties are identical and can be treated
as a single species), modulated by the large value of solar mixing
angle, $\theta_{12}$, well determined by solar and KamLAND
data~\cite{Maltoni:2004ei,Schwetz:2008er}.
By performing a suitable simulation of the high-statistics SN data it
was shown in Refs.~\cite{Minakata:2001cd,Barger:2001yx} how one can
determine the parameters of the original neutrino spectra,
$\bar{\nu}_e$ and $\nu_x$ in terms of the $\bar\nu_e$ signal detected
in  inverse beta decay.
However, this result has been obtained under the assumption that we
know the parameter which describes the ``pinching'' of the SN neutrino
fluxes.  (See Sec.~\ref{subsec:basic} for the pinching.)

In this paper, we extend our previous work in \cite{Minakata:2001cd}
by including the pinching of the SN neutrino spectra as fit
parameters\footnote{A study of the relevance of this parameter in the
  analysis of the neutrino signal from SN1987A can be found for
  instance in Refs.~\cite{janka1989net,Mirizzi:2005tg}.}.
In the extended framework, regrettably enough, we face with an
important difficulty in reconstructing the original neutrino spectra.
Most significantly, we find that there exists a persistent continuous
degeneracy in the flavor-dependent determination of the luminosity and
the spectra of SN neutrinos.

\begin{figure}[bhtp]
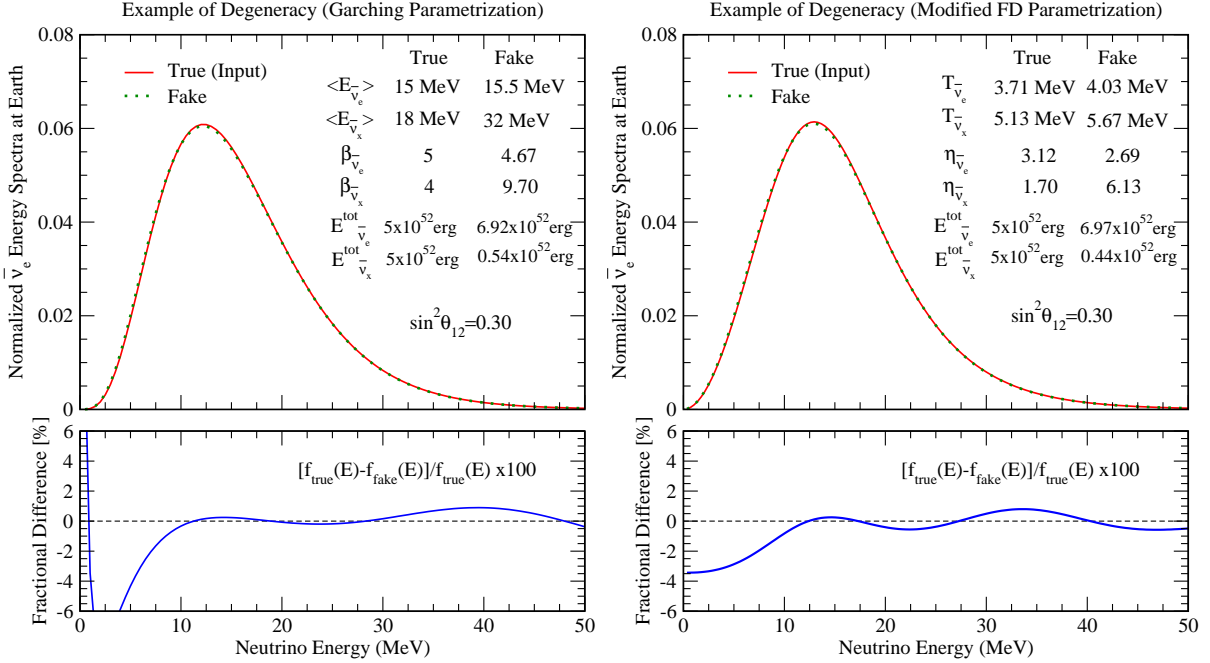

\vglue 0.2cm
\begin{center}
\hglue -0.2cm
\includegraphics[width=0.48\textwidth]{fig1a.eps}
\includegraphics[width=0.48\textwidth]{fig1b.eps}
\end{center}
\vglue -0.2cm
\caption{
Examples of degenerate determination of SN astrophysical parameters. 
The two degenerate neutrino energy spectra at Earth (upper panels) as
well as their fractional difference (lower panels) resulting from
neutrino fluxes with two different energy spectra at the SN core and
the indicated parameter sets. The left and right panels correspond to
the Garching parametrization and the modified Fermi-Dirac
parametrization, respectively. Input values are the same 
ones shown in Table ~\ref{tab:input}. 
While it is not clear from the legend in the right panel, 
we note that the both true and fake values of average 
energies for the Fermi-Dirac parametrization give the same 
values as in the case of Garching one. 
Normal neutrino mass hierarchy has been assumed. 
Notations of neutrino species and the definitions of the
parameters involved are given in Sec.~\ref{sec:SNnu}.  }
\label{fig:example}
\end{figure}

In order to make our point and to indicate how serious the degeneracy
problem is, we show in Fig.~\ref{fig:example} examples of two
degenerate $\bar{\nu}_e$ energy spectra at the Earth (upper panels) as
well as their difference divided by the true spectrum or fractional
difference (lower panels) for two different parametrizations of the SN
neutrino flux.
It is remarkable that, despite the large difference in the primary
neutrino spectra at the SN core particularly for $\nu_x$, 
the true and fake spectra agree with
each other within better than 1\% level over a wide energy range.
It is evident that the degeneracy is present already in the bare
neutrino fluxes reaching the detector (before taking into account
experimental uncertainties such as energy resolution) and it looks so
perfect that it is unlikely to be resolved even with the extremely
high statistics expected in a Megaton
detector~\cite{Okumura:2007,UNO,MEMPHYS}.
Hopefully, our negative result will stimulate further studies towards
a full diagnosis of the SN core by neutrino observations.  These
include a better theoretical understanding of the neutrino fluxes
formation, as well as an optimization of the information provided by
the different reactions in the neutrino detectors.

The plan of the paper is as follows. In Sec.~\ref{sec:SNnu} we
describe the basic features of supernova neutrinos, including the
different parametrizations of the initial neutrino spectra usually
considered in the literature, as well as the effect of neutrino flavor
conversion before reaching the detector.
In Sec.~\ref{sec:assumption} we will specify the assumptions used in
our analysis. Those include the particular values characterizing the
initial neutrino fluxes, the details of the detector, and the neutrino
mass scheme. In Sec.~\ref{sec:degeneracy} we demonstrate the existence
of a continuous degeneracy in the determination of the astrophysical
parameters. We discuss the robustness of this finding, which results
neither from the particular parametrization taken, nor from a
particular set of SN parameters.  Finally, in Sec.~\ref{sec:solveDG}
we speculate possible ways that could help overcoming the degeneracy
problem.

\section{Supernova neutrinos}
\label{sec:SNnu}

\subsection{Basic features of neutrino spectra from supernova}
\label{subsec:basic}

To a crude approximation the proto-neutron star is a black-body source
for neutrinos of all flavors.  For the case of $\nu_e$ and $\bar\nu_e$
the dominant reactions are the charged-current (CC) interactions with
nucleons $e^- + p \leftrightarrow \nu_e + n$ and $e^+ + n
\leftrightarrow \bar\nu_e + p$.
The other flavors, $\nu_{\mu}$, $\bar\nu_{\mu}$, $\nu_{\tau}$,
$\bar\nu_{\tau}$ which in this paper we shall collectively denote by
$\nu_x$, interact with the surrounding matter via neutral current (NC)
interactions, e.g. Bremsstrahlung, neutrino-pair annihilation or
neutrino-nucleon scattering.  These processes keep $\nu_e$,
$\bar\nu_e$ and $\nu_{x}$ in local thermal equilibrium up to the radii
where these reactions become inefficient (neutrinosphere).  Beyond
these radii neutrinos freely stream.
Taking into account the hierarchy in the cross sections, $\sigma_{CC}
> \sigma_{NC}$ as well as the richer neutron composition than protons,
one expects the average neutrinosphere radii of $\nu_{x}$,
$\bar\nu_e$, and $\nu_e$ to obey $r_x < r_{\bar{e}} < r_e$, so that
$\nu_{x}$ ($\nu_e$) decouples at the highest (lowest) temperature.
This translates to an ordering of the average energies of SN neutrinos
$\vev{E_{\nu_{e}}} < \vev{E_{\bar\nu_{e}}} < \vev{E_{\nu_{x}}}$, the
exact degree of difference still under
debate~\cite{Raffelt:2001kv,Buras:2002wt,Raffelt:2003en}.
See, for example, \cite{Keil:2002in,Mezzacappa:2005ju} for more 
about physics in the proto-neutron star. 

The location of the neutrinospheres does not only depend on the
neutrino flavor but also on its energy. The energy dependence of the
 cross sections of the processes involved makes neutrinos with
different energies decouple from the proto-neutron star at different
radii and therefore different local temperatures. For this reason the
spectrum of the neutrinos leaving the star does not present a thermal
distribution.  The possibility of reconstructing the flux parameters
of the three effective flavors from observation would lead to a
``neutrino imaging'' of the proto-neutron star.

There are different ways to characterize the non-thermal spectra of
the neutrino fluxes emerging from the SN. Among them there are two
parametrizations that have been extensively used in the literature. 
One is the 
Fermi-Dirac distribution motivated by the equilibrium
distribution of neutrinos inside the star~\cite{janka1989net}
\begin{equation}
  F^0_{\nu_\alpha}(E) = 
  \frac{\Phi_{\nu_\alpha} 
  }{T_{\nu_\alpha}^3f_2(\eta_{\nu_\alpha})} 
  \frac{E^2}{e^{E/T_{\nu_\alpha}-\eta_{\nu_\alpha}}+1}, 
\label{eq:flux-FD}
\end{equation}
where $E$ is the neutrino energy, and $T_{\nu_\alpha}$ and
$\eta_{\nu_\alpha}$ denote an effective temperature and degeneracy
parameter (chemical potential), respectively.  The distribution is
normalized so that $\Phi_{\nu_\alpha}$ stands for the total number of
$\nu_\alpha$ emitted.
The function $f_n(\eta_{\nu_\alpha})$ is defined as 
\begin{equation}
f_n(\eta_{\nu_\alpha}) \equiv \int_0^\infty 
\frac{x^n}{{\rm e}^{x-\eta_{\nu_\alpha}}+1}{\rm d}x \,.
\end{equation}
The mean energy and the total energy released are 
consequently 
$\vev{E_{\nu_\alpha}} = 
\left[ f_3(\eta_{\nu_\alpha}) / f_2(\eta_{\nu_\alpha}) \right]
T_{\nu_\alpha}$ and 
 $E_{\nu_\alpha}^{\rm tot} = \Phi_{\nu_\alpha}
\vev{E_{\nu_\alpha}}$, respectively.
Note that it must be understood that $\Phi_{\nu_x}$ does not refer
to the sum of the flux of the non-electron species (despite that we treat
them as a single species) but the individual one as follows,
\begin{equation}
\Phi_{\nu_x} = \Phi_{\nu_\mu} =  \Phi_{\bar{\nu}_\mu} = \Phi_{\nu_\tau} =
\Phi_{\bar{\nu}_\tau},
\end{equation}
and so as for  $E_{\nu_x}^{\rm tot}$, throughout this paper.

\begin{figure}[ht!]
  \begin{center}
    \includegraphics[width=0.50\textwidth,angle=0]{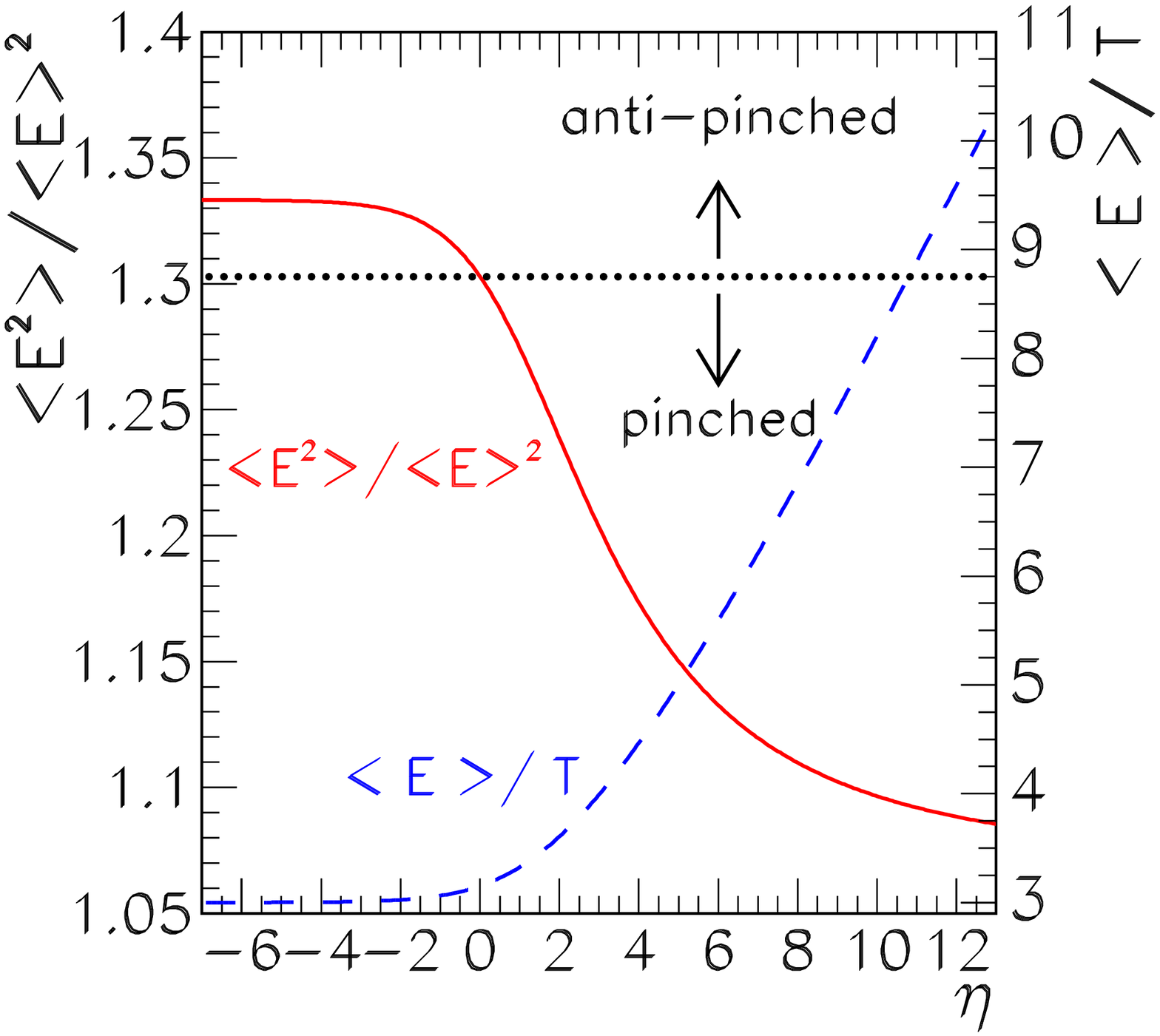}
    \hspace*{0.1cm}
    \includegraphics[width=0.48\textwidth,angle=0]{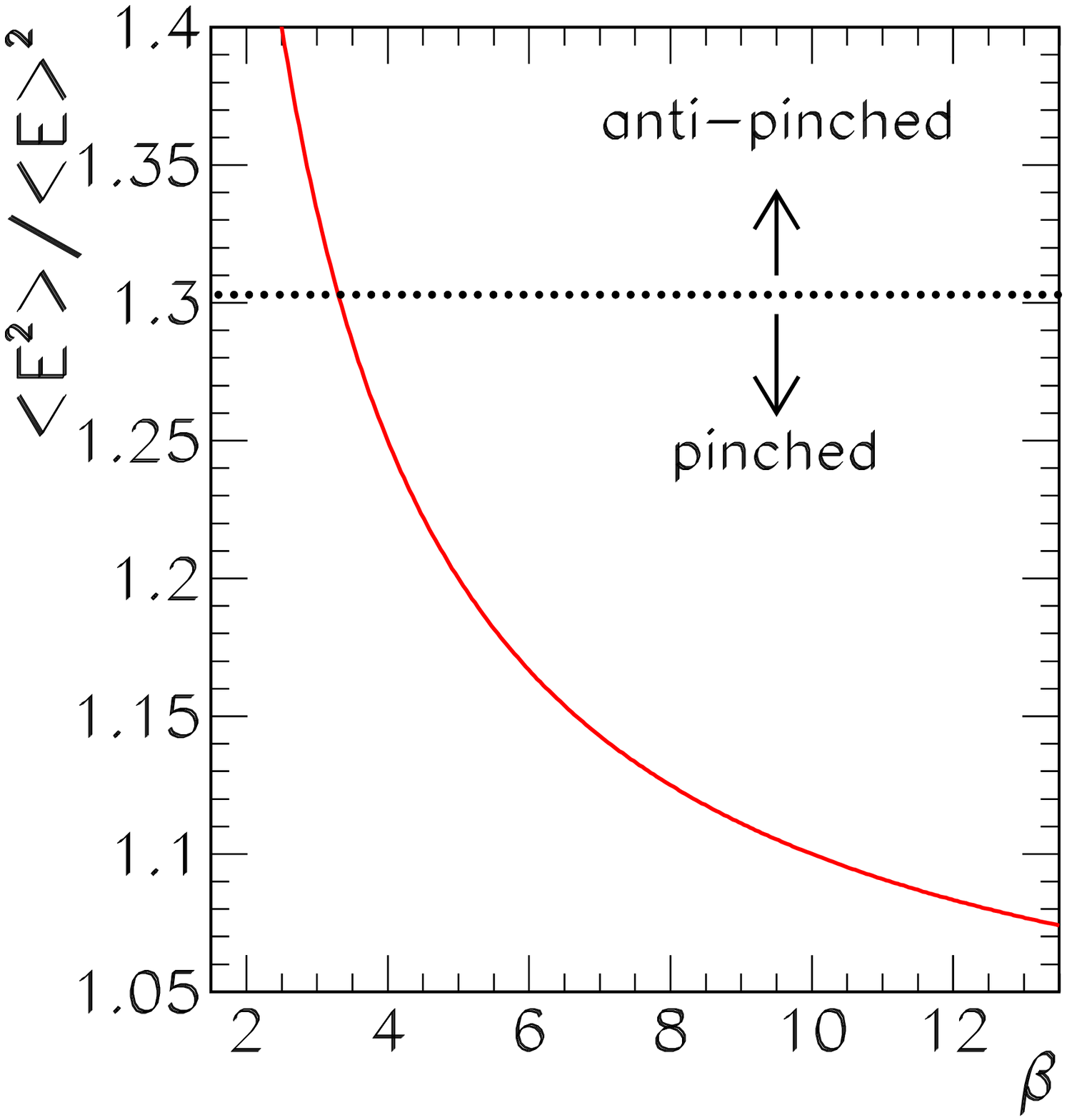}
  \end{center}
\vglue -0.8cm
\caption{The Left panel gives the pinching parameter $p \equiv
  \vev{E^2} / \vev{E}^2$ (left label) and the ratio $\vev{E} / T $
  (right label) for the Fermi-Dirac distribution represented as a
  function of $\eta$ by the solid and dashed curves, respectively. The
  critical value of $p$ which separates between a pinched and an
  anti-pinched distribution is indicated by the dotted line. The right
  panel gives the pinching parameter for the Garching parametrization
  as a function of $\beta$.}
  \label{fig:ap}
\end{figure}

A way to determine how much a spectrum deviates from being thermal
is to use the pinching parameter~\cite{Raffelt:2001kv} defined 
as the ratio of the first two moments

\begin{equation}
p \equiv \frac{\vev{E^2}}{\vev{E}^2} \,.
\end{equation}
A spectrum that is thermal up to its second moment has $p=p_{{\rm
    FD},\eta=0}\approx 1.3029$, while $p<p_{{\rm FD},\eta=0}$ implies
a pinched spectrum (high- and low-energy parts of the spectrum
relatively suppressed) and $p>p_{{\rm FD},\eta=0}$ is an anti-pinched
spectrum (high- and low-energy parts of the spectrum enhanced).  For
the Fermi-Dirac distribution with an arbitrary $\eta$ the pinching
parameter $p$ is related to $\eta$ as $$p = [ f_4(\eta)f_2(\eta) / 
  f_3^2(\eta) ] $$.
In the left panel of Fig.~\ref{fig:ap} we show in solid red lines the
explicit dependence of the pinching parameter on the effective
degeneracy parameter $\eta$. 

The curve becomes flat at negative $\eta$, which reflects the fact that 
the function tends to the Maxwell-Boltzmann spectrum at 
$\eta\rightarrow - \infty$, which does not differ much from the 
Fermi-Dirac one, $\eta=0$. 
In dashed lines we present the strong dependence of 
$\vev{E} / T$ on $\eta$ for pinched distributions.

\begin{figure}[ht!]
  \begin{center}
    \includegraphics[width=0.50\textwidth,angle=0]{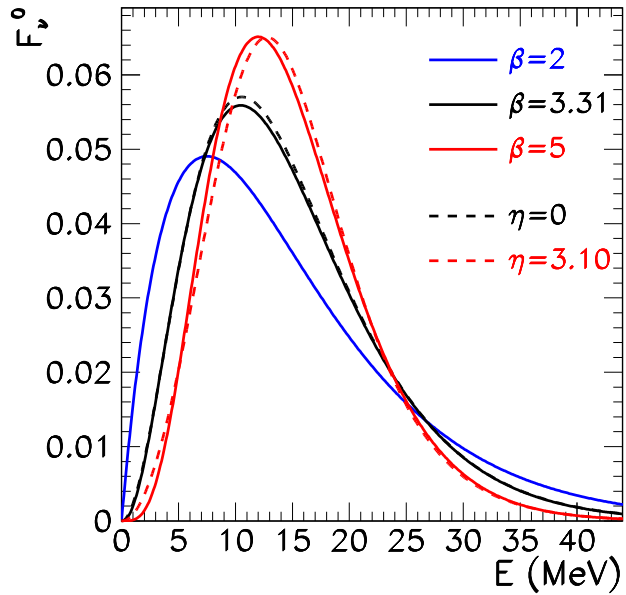}
  \end{center}
\vglue -0.8cm
\caption{The solid curves show examples of anti-pinched ($\beta=2$,
  shown in blue), un-pinched ($\beta=3.31$, in black), and pinched
  ($\beta=5$, in red) Garching distributions normalized to unity.  The
  mean energy has been set to $15$~MeV.  The corresponding Fermi-Dirac
  distributions for the same value of $\vev{E}$ and $p$ are also shown
  by the dashed lines.  Note that there is essentially no anti-pinched
  Fermi-Dirac distribution for $\beta=2$ because of the saturation
  property of $p$ shown in Fig.~\ref{fig:ap}.  }
  \label{fig:ap2}
\end{figure}

A second parametrization using the form of power times exponential has
been recently introduced by the Garching group as a better
parametrization of their simulation results
\cite{Keil:2002in,Keil:2003sw,Buras:2002wt}
\begin{eqnarray}
  F^0_{\nu_\alpha}(E) =
  \frac{\Phi_{\nu_\alpha}}{\vev{E_{\nu_\alpha}}}\,
  \frac{\beta_{\nu_\alpha}^{\beta_{\nu_\alpha}}}{\Gamma(\beta_{\nu_\alpha})}  
  \left(\frac{E}{\vev{E_{\nu_\alpha}}}\right)^{\beta_{\nu_\alpha}-1} 
  \exp\left(-\beta_{\nu_\alpha}\frac{E}{\vev{E_{\nu_\alpha}}}\right) \,.
\label{eq:flux-Gal}
\end{eqnarray}
In this case the parameters characterizing the distribution function
are the total number of $\nu_\alpha$ emitted, $\Phi_{\nu_\alpha}$, the
mean energy, $\vev{E_{\nu_\alpha}}$, and a parameter
$\beta_{\nu_\alpha}$ which is related to the pinching parameter via
$$p= (\beta_{\nu_\alpha} + 1)/\beta_{\nu_\alpha}.$$
In the right panel of Fig.~\ref{fig:ap} we show in solid red lines the
dependence of $p$ on $\beta_{\nu_\alpha}$.  In contrast to the
Fermi-Dirac parametrization there is no asymptotic limit for the width
$\vev{E^2} / \vev{E}^2$, hence this parametrization can reproduce
better anti-pinched distributions.
In Fig.~\ref{fig:ap2} we illustrate the effect of the pinching on
three spectra with different $\beta$'s. One can see how the pinched
spectrum is suppressed at low and high energies with respect to the
non-pinched one. This in turn shows a similar suppression in
comparison with the anti-pinched spectrum.
We observe in the same figure how this behavior holds also for the
case of the Fermi-Dirac distributions. 
The results shown above agree with 
those found in \cite{Keil:2002in} where the similarities 
as well as differences of these 2 parametrizations were explored.

\subsection{Flavor conversion of supernova neutrinos}
\label{subsec:conversion}

The dynamics of neutrino flavor conversion in a typical iron-core SN
can be factorized into two different parts~\footnote{
There are also particular scenarios with very shallow electron density
profiles, like the early stages of O-Ne-Mg core
SNe~\cite{Janka:2007di}, where the dynamics in the inner and outer
regions can not be decoupled. In this case collective effects and MSW
resonances are not clearly separated, what can lead to interference
effects both in normal and inverted mass hierarchy~\cite{Duan:2007sh}.
In the following, though, we will not consider this particular case.
}: the propagation through the inner layers, where the high neutrino
density can lead to non-linear collective
effects~\cite{Duan:2005cp,Duan:2006an,Duan:2006jv,Hannestad:2006nj,Raffelt:2007cb,EstebanPretel:2007ec,Duan:2007bt,Fogli:2007bk,Duan:2007sh,EstebanPretel:2007yq,Dasgupta:2007ws},
and the evolution in the outer layers.
Right above the neutrinosphere the neutrino density is so high that
neutrino-neutrino interactions must be taken into account. The
presence of neutrino self-interactions potentially give rise to
collective phenomena on the neutrino propagation. These include
synchronization, pair conversions of the kind
$\nu_e\bar\nu_e\rightarrow \nu_x\bar\nu_x$ during the so-called
bipolar regime, and spectral split. These effects, though, arise only
in the case of inverted neutrino mass hierarchy.

In the outer layers the neutrino-neutrino interactions can be
neglected and therefore it is enough to consider the
Mikheev-Smirnov-Wolfenstein (MSW) effects induced on neutrinos by
background matter~\cite{Wolfenstein:1977ue,Mikheev:1986gs}.
The neutrino flavor evolution can be then described in terms
of two independent two-level crossings associated to the atmospheric
and solar mass squared splittings~\cite{Kuo:1989qe,Dighe:1999bi}.  Let
us denote the former (latter) the H (L) level crossing.  The relation
is simplified by the fact that the L level crossing is adiabatic,
given the confirmation of the large mixing angle solution by the
KamLAND data.

Yet, there are still uncertainties which arise due to the two
remaining unknowns, the neutrino mass hierarchy and the value of
$\theta_{13}$. The former leads to the freedom of the H level crossing
in either the neutrino or the anti-neutrino channels depending upon the
normal or the inverted mass hierarchies, respectively, which in
principle can be used to discriminate the hierarchy
\cite{Dighe:1999bi,Minakata:2000rx,Takahashi:2002cm,Lunardini:2003eh,Dighe:2003be,Dighe:2003jg,Dighe:2003vm,Barger:2005it,Skadhauge:2006su}.
The latter allows the possibilities of the adiabatic or the 
non-adiabatic H level crossing depending upon 
$\sin^2\theta_{13} \gsim 10^{-3}$ or  $\sin^2\theta_{13} \lsim 10^{-5}$.

\section{ Assumptions }
\label{sec:assumption}

\subsection{ SN flux model }
\label{subsec:flux}

We employ a SN model-dependent fitting method to reconstruct SN
neutrino fluxes at the core.  We trust global features of the SN
neutrino fluxes at the core obtained by detailed simulations and
parametrize them by simple functions.  We adopt two choices, namely,
the traditional Fermi-Dirac form Eq.~(\ref{eq:flux-FD}) and the
Garching parametrization in Eq.~(\ref{eq:flux-Gal}).
Both parametrizations contain three parameters, ($\Phi_{\nu_\alpha},
~\vev{E_{\nu_\alpha}}, ~\beta_{\nu_\alpha}$) and ($\Phi_{\nu_\alpha},
~T_{\nu_\alpha}, ~\eta_{\nu_\alpha}$) for the Garching and the
Fermi-Dirac distributions, respectively.
In previous
works~\cite{Barger:2001yx,Minakata:2001cd,GilBotella:2004bv,Skadhauge:2006su}
the pinching parameters were always fixed and thus assumed to be known
in advance.
The inclusion of $\beta_{\nu_\alpha}$ ($\eta_{\nu_\alpha}$) parameter
as a fit parameter constitutes therefore a new feature of the present
analysis.  We feel that this inclusion is essential because the
pinching parameter represents the effect of departure from local
thermal equilibrium and would reflect the environmental condition of
matter around the neutrinosphere.  Hence, it should be determined by
observations.
The values of initial parameters used to generate SN neutrino ``data'' 
are summarized in Table~\ref{tab:input}. 
We assume that SN neutrino spectra are time-independent during the
burst.
Although this may be too idealized, it is a conservative assumption in
the sense that relaxing this it would further complicate the task of
resolving the degeneracy.

\begin{table}
\begin{center}
\begin{tabular}{llccc}
  \hline
  Flux model &  average energy ($\bar\nu_{e}$) & average energy ($\nu_{x}$) &  pinching ($\bar\nu_{e}$) & pinching ($\nu_{x}$)  \\ 
  \hline
  Garching &  $\vev{E_{\bar{\nu}_{e}}} = 15$ MeV  &  $\vev{E_{\nu_{x}}} = 18$ MeV & $\beta = 5$ & $\beta = 4$ \\
Fermi-Dirac & $T_{\bar\nu_e} = 3.71$~MeV & $T_{\nu_x}= 5.13$~MeV &  $\eta = 3.12$ & $\eta = 1.70$ \\
\hline
\end{tabular}
\caption{Reference input values of the Garching and the Fermi-Dirac
  flux parameters used in this work. The total energies and the width
  $\vev{E^2} / \vev{E}^2$ of $\bar\nu_{e}$ and $\nu_{x}$ are taken as
  follows: $E^{\rm tot}_{\bar\nu_e} = E^{\rm tot}_{\nu_x} = 5\times
  10^{52}~{\rm erg}$, $p_{\bar\nu_e} = \vev{E_{\bar\nu_e}^2} /
  \vev{E_{\bar\nu_e}}^2= 1.2$, and $p_{\bar\nu_x} = \vev{E_{\nu_x}^2}
  / \vev{E_{\nu_x}}^2= 1.25$.  The flux $\Phi_{\nu_\alpha}$ is determined by the relation
$E_{\nu_\alpha}^{\rm tot} = \Phi_{\nu_\alpha} \langle E_{\nu_\alpha}
\rangle$.  The temperatures and the $\eta_{\nu_\alpha}$ of the
Fermi-Dirac distribution are chosen to reproduce the same average
energies listed in the upper row and the same $p$.
}
\label{tab:input}
\end{center} 
\end{table} 

\subsection{ Detector and analysis method}
\label{subsec:detector}

In this paper, we focus on water Cherenkov detectors as they are most
likely the ones to run sufficiently long so as to watch galactic
supernovae over long enough time scales. In particular we consider
Hyper-Kamiokande \cite{Okumura:2007}, and take the fiducial mass of
720 kton assuming that the whole inner volume is available for SN
neutrinos.  For other similar projects see
e.g. Refs.~\cite{UNO,MEMPHYS}.

In this kind of detectors several reactions contribute to the SN
neutrino signal: inverse beta decay, elastic scattering off
electrons, and CC and NC with oxygen. 
Nevertheless, 
inverse beta decay is indeed the dominant reaction, which would yield
$\sim 2 \times 10^5$ events in Hyper-Kamiokande for a SN at 10 kpc
from the Earth.  For the cross section see
Ref.~\cite{Vogel:1999zy,Strumia:2003zx}.  The expected event numbers
from $\nu_{e}$ and $\nu_{x}$ elastic scattering and $\nu_e$ absorption
by Oxygen are more than an order of magnitude below~\cite{Burrows:1991kf}.

In this paper, thus, we restrict our analysis into a unique observable,
the positron energy spectra produced by the $\bar{\nu}_{e}$ absorption
reaction on protons, $\bar{\nu}_{e} + p \to e^+ + n$.
For this reaction only six parameters are relevant:
$E^{\rm tot}_{\bar\nu_e}$, $\vev{E_{\bar\nu_e}}$, $p_{\bar\nu_e}
\equiv \vev{E^2_{\bar\nu_e}} / \vev{ E_{\bar\nu_e}}^2$,
$E^{\rm tot}_{\nu_x} $, $\vev{E_{\nu_x}}$, and $p_{\nu_x} \equiv
\vev{E^2_{\nu_x}} / \vev{ E_{\nu_x} }^2$.  
We assume the SN at a distance of 10 kpc from the Earth and generate
the SN neutrino flux data by assuming the seed parameters listed in
Table~\ref{tab:input}.  The data are subsequently fitted using the
same parametrization assumed to generate it, except for results shown
in Fig.~\ref{fig:FD-Keil}.
Motivated by our current understanding of the composition of the
proto-neutron star we also assume that the ratio~\footnote{To extract
  the {\em real degeneracy} we want to 
    eliminate a trivial degeneracy which inevitably comes in into such
    fitting procedure.  In view of Eq.~(\ref{eq:F_A}) it is clear that
    for any solution for the six parameters labeled as (a), there is
    another solution labeled as (b) in which
    $\beta_{\bar\nu_e}^{(b)}$, $\vev{ E_{\bar\nu_e}}^{(b)}$ are
    replaced by $\beta_{\nu_x}^{(a)}$, and $\vev{ E_{\nu_x}}^{(a)}$,
    and vice versa, together with $\Phi_{\bar\nu_e}^{(b)} = \tan^2
    \theta_{12} \Phi_{\bar\nu_e}^{(a)}$, $\Phi_{\nu_x}^{(b)} = \cot^2
    \theta_{12} \Phi_{\nu_x}^{(a)}$.
    In these expressions, we have used the approximation $s_{13} \ll
    1$ for simplicity.  We have explicitly verified that the exchange
    degeneracy solution can be removed by imposing the former
    condition.  } $\tau\equiv \vev{ E_{\nu_x}} / \vev{ E_{\bar\nu_e}}
  > 1$.
We perform a standard $\chi^2$ analysis including only the statistical
error.  The energy bins used are chosen as: 70 bins for $5 \text{MeV}
\leq E_e \leq 40 \text{MeV} $, 10 bins for $40 \text{MeV} \leq E_e
\leq 50 \text{MeV} $, 3 bins for $50 \text{MeV} \leq E_e \leq 56
\text{MeV} $, 1 bin for $56 \text{MeV} \leq E_e \leq 60 \text{MeV} $,
and the highest energy bin for $60 \text{MeV} \leq E_e \leq 100
\text{MeV} $ (total 85 bins), where $E_e$ stands for the measured
energy of the positrons.
The energy resolution is taken into account with a Gaussian function
with width $\sigma_{\text{res}} = 0.47 \sqrt{E_e/\text{MeV}}$~MeV,
following Ref.~\cite{Hosaka:2005um}.
The results are represented in terms of $E^{\rm tot}_{\nu_\alpha},
~\vev{ E_{\nu_\alpha}}$ and $\vev{ E^2_{\nu_\alpha}} /
\vev{E_{\nu_\alpha}}^2$.
The 2 (3) $\sigma$ CL allowed regions are determined by the condition, 
\begin{equation}
\Delta \chi^2 = \chi^2 - \chi^2_{\text{min}} < 6.18\  (11.83), 
\end{equation}
for 2 degrees of freedom. 

\subsection{Neutrino mass hierarchy and validity of our ansatz of SN neutrino flux spectra}
\label{subsec:scheme}

In this paper we assume that SN neutrino spectra at the surface of the
progenitor star are given by the flavor transformed ones of the
initial spectra of either the Garching or the Fermi-Dirac type.

In order to illustrate the possibility of inferring the original
neutrino spectra we consider the case with normal mass hierarchy.
Ignoring the Earth matter effect, the relationships between the
$\bar\nu_e$ flux at the core of SN and the one at terrestrial
detectors is as follows:
\begin{equation}
F_{\bar\nu_e} =
c^2_{12} c^2_{13} F_{\bar\nu_e}^0 + 
(1- c^2_{12} c^2_{13}) F_{\nu_x}^0 
\approx c^2_{12} F_{\bar\nu_e}^0 + s^2_{12} F_{\nu_x}^0 \,, 
\label{eq:F_A}
\end{equation}
where the expressions $s^2_{12},~c^2_{12},~s^2_{13}$, and $~c^2_{13}$
stand for $\sin^2\theta_{12},~\cos^2\theta_{12},~\sin^2\theta_{13}$,
and $\cos^2\theta_{13}$, respectively. 
The observed $\bar\nu_e$ spectra is then a superposition of the
original $\bar\nu_{e}$ and $\nu_{x}$ spectra. The coefficients of the
composition depend basically on the value of $\theta_{12}$, which we
will assume to be
$\sin^2\theta_{12}=0.3$~\cite{Maltoni:2004ei,Schwetz:2008er}.
Therefore the goal of the analysis is to disentangle the two different
components present in the observed $\bar\nu_e$ spectrum.

The validity of our analysis would be affected if any mechanisms are
operational inside or outside SN core that invalidate the above
ansatz. These include basically three possibilities. 
The first one is a possible unknown time dependence of the parameters
characterizing the initial neutrino fluxes. In this case one can
always carry out the analysis for considering the positron spectra at
different time bins. The main consequence of this effect will be an
increase of the statistical errors but the main features would remain.

The second aspect that could affect our analysis is the presence of
the shock waves propagating within the supernova. In several
works~\cite{Schirato:2002tg,Fogli:2003dw,Tomas:2004gr,Fogli:2004ff}
it has been discussed how the presence of these discontinuities could
introduce an energy and time modulation in the survival probabilities.
However, the $\bar\nu_e$ flux is only affected in the case of inverted
mass hierarchy and ``large'' $\theta_{13}$, $\sin^2\theta_{13}\gtrsim
10^{-4}$. 
The formation of shock waves could additionally lead to turbulent
density fluctuations, causing a neutrino flavor
depolarization as  discussed in ~\cite{Fogli:2006xy} 
and ~\cite{Friedland:2006ta} 
where the former considered 
the $\delta$-correlated density fluctuation whereas the latter 
considered the Kolmogorov-type turbulence implied by realistic 
SN simulations.  
The characteristic signal due to the shock wave, which was
originally considered visible ignoring the effects of turbulence
and/or density fluctuations, tend to be washed out 
by such turbulent density fluctuations ~\cite{Fogli:2006xy,Friedland:2006ta}. 
See also ~\cite{Kneller:2007kg,Dasgupta:2005wn} 
for effects related to the shock waves.
We note, however, that as long as the $\bar\nu_e$ flux 
(at Earth) is concerned,  these effects would be significant 
only in the case of inverted mass  hierarchy and not so 
small value of $\theta_{13}$, 
$\sin^2\theta_{13}\gtrsim O(10^{-3})$.

%

The third case where our analysis would not apply would be in the
presence of decoherence. This can be due to the multi-angle nature of
the neutrino self-interaction, and can be present for both normal and
inverted mass hierarchies. Nevertheless, it is expected to
significantly affect the neutrino propagation only for very similar
$\nu_e$ and $\bar\nu_e$ spectra~\cite{EstebanPretel:2007ec}, which
does not seem to be the case according to the SN
simulations~\cite{Keil:2003sw}. Therefore we will neglect it.

Last, but not least, the possible existence of non-standard neutrino
interactions, apart from affecting neutrino propagation through the SN
envelope, could induce resonant conversions in the most deleptonised
inner layers~\cite{Valle:1987gv} with possibly dramatic effects in a
Megaton water Cherenkov detector~\cite{EstebanPretel:2007yu}. This
possibility will also not be considered here.

Before closing this section let us mention that a similar analysis
could be performed in the case of inverted mass hierarchy. However, in
this case one should take into account the possible modulations
induced by the shock wave passage, as well as by collective effects
such as the spectral split in $\bar\nu_e$~\cite{Fogli:2007bk} or the
effect of the second-order difference between the $\nu_\mu$ and
$\nu_\tau$ refractive index~\cite{EstebanPretel:2007yq}.  This
comparative study lies however beyond the scope of this paper.

\section{Degeneracy in fitted parameters}
\label{sec:degeneracy}

\subsection{Continuous degeneracy in the fit parameters}  
\label{subsec:degeneracy}

\begin{figure}[h]
  \begin{center}
    \includegraphics[width=0.33\textwidth]{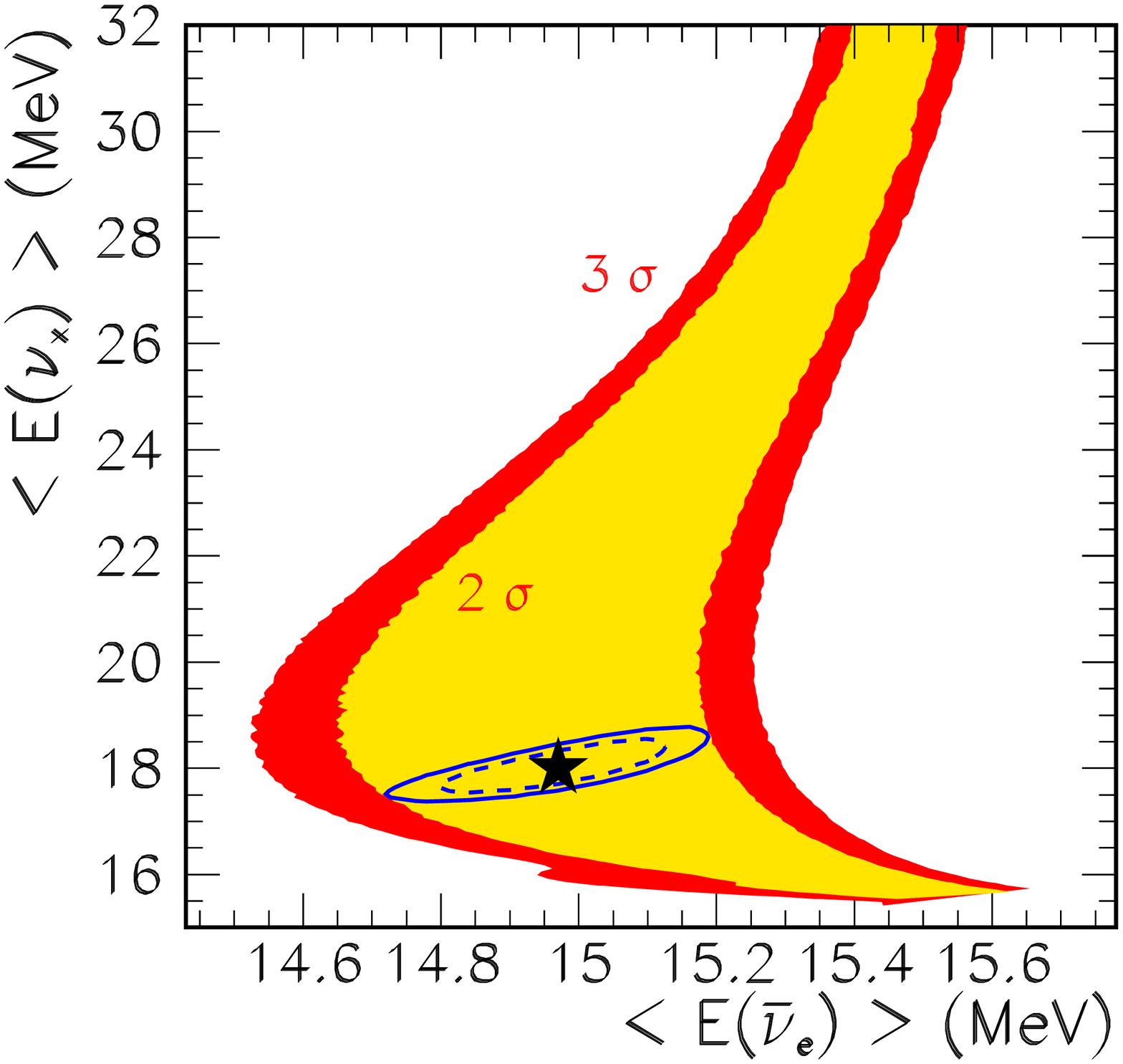}
    \includegraphics[width=0.32\textwidth]{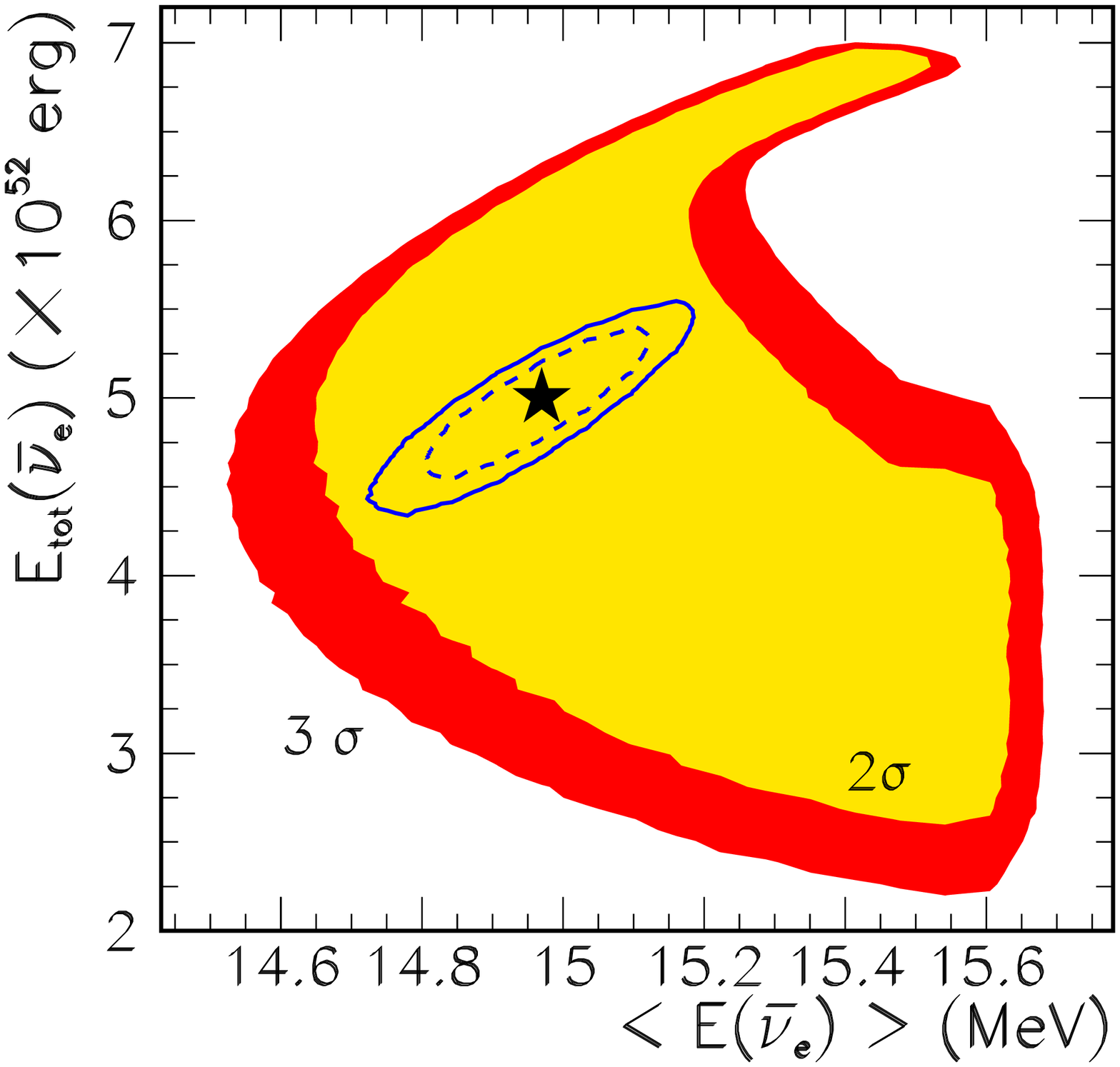}
    \includegraphics[width=0.32\textwidth]{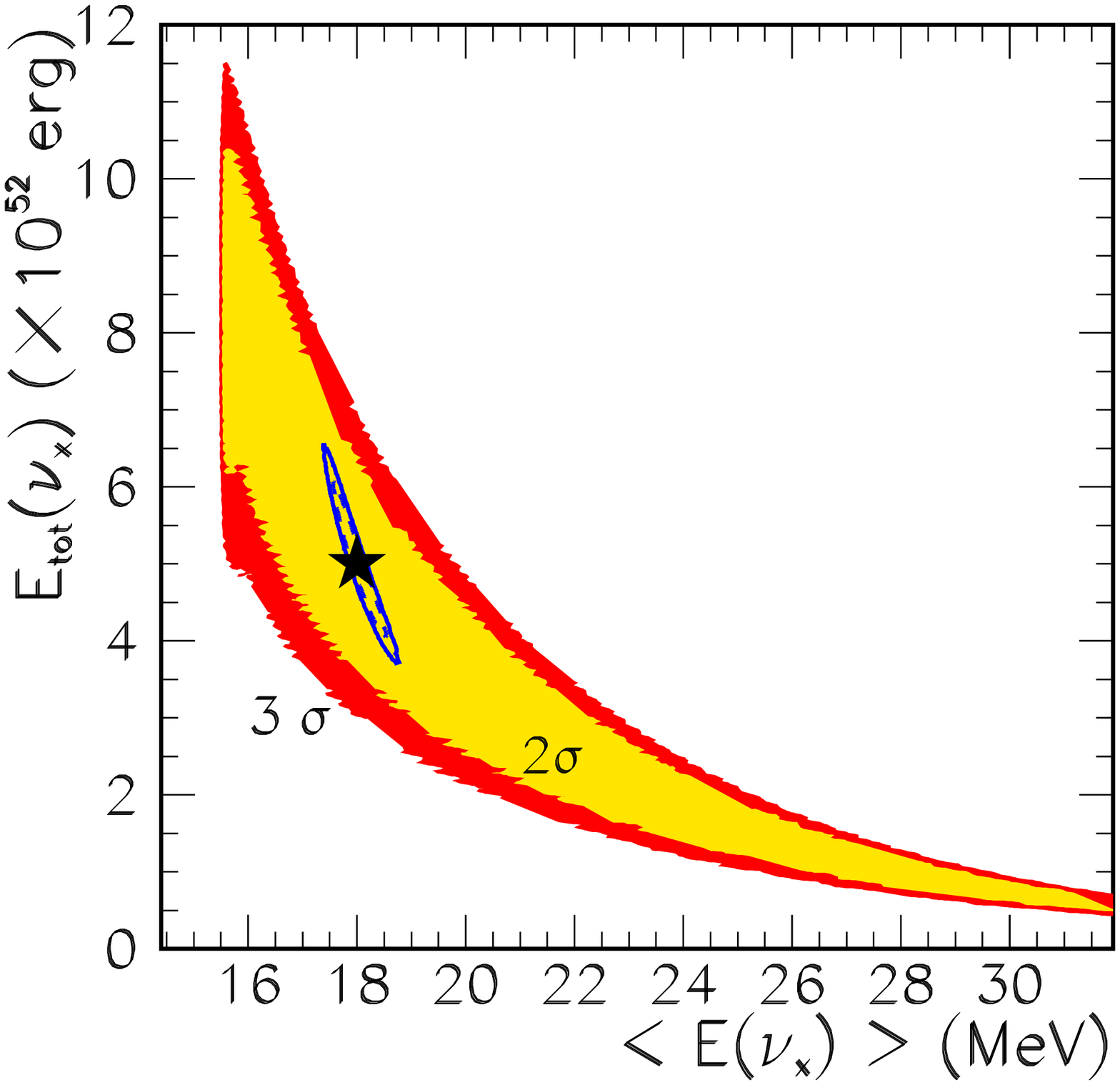}
  \end{center}
\vglue -0.8cm 
\caption{Comparison between the 2$\sigma$ CL and 3$\sigma$ CL determination 
of the astrophysical parameters for the cases of fixed and free pinching 
parameters, displayed in terms of 
$\langle E_{\bar\nu_e} \rangle - \langle E_{\nu_{x}} \rangle $ (left panel), 
$\langle E_{\bar\nu_e} \rangle - E^{\rm tot}_{\bar\nu_e}$ (middle panel), and 
$\langle  E_{\nu_x} \rangle - E^{\rm tot}_{\nu_x}$ (right panel).
  The solid and the dashed ellipses indicate the cases in which 
$\beta_{\bar{\nu}_{e}}$ and $\beta_{\nu_{x} }$ are fixed to $\beta_{\bar{\nu}_{e}}=5$ and
  $\beta_{\nu_{x} }=4$.  The star represents the best-fit point,
  which, in this case, coincides with the input values. The constraint
  $\tau\equiv \langle E_{\nu_x}\rangle / \langle E_{\bar\nu_e} \rangle
  \ge 1$ has been imposed.
  The corresponding at 2$\sigma$ CL and 3$\sigma$ CL determinations
  for the case of free pinching parameters are denoted by the
  (dark/red) and (light/yellow) regions.  See the text and
  Tab.~\ref{tab:input} for more details.}
\label{fig:degeneracy1} 
\end{figure}

Let us start with the analysis with the Garching parametrization of
the SN $\nu$ fluxes assuming the values for the initial $\nu$ spectra
given in Tab.~\ref{tab:input}.  To understand the effect of varying
pinching parameters let us first assume that the $\beta_{\nu_\alpha}$
parameters are known. 
In Fig.~\ref{fig:degeneracy1} we represent by the dashed (solid) blue
ellipses the regions allowed at $2\sigma$ ($3\sigma$) CL in the space
spanned by $\langle E_{\bar\nu_e} \rangle - \langle E_{\nu_{x}}
\rangle $ (left panel), $\langle E_{\bar\nu_e} \rangle - E^{\rm
  tot}_{\bar\nu_e}$ (middle panel), and $\langle E_{\nu_x} \rangle -
E^{\rm tot}_{\nu_x}$ (right panel).
In each panel, the best fit point is also indicated by a star, which
of course reproduces the input value.  It can be seen how we can
determine $\langle E_{\bar\nu_e} \rangle$, $\langle E_{\nu_x}
\rangle$, $E^{\rm tot}_{\bar\nu_e}$, and $E^{\rm tot}_{\nu_x}$ with an
accuracy of roughly 2\%, 4\%, 15\%, and 30\%, respectively, at
3$\sigma$ CL.  These results are in good agreement with our previous
work \cite{Minakata:2001cd} apart from small differences due to
different assumptions on the initial spectra and the detector (now
with slightly smaller volume).  It is remarkable that the
non-vanishing mixing angle $\theta_{12}$ allows us to obtain
information about $\nu_x$ flux parameters event though only
$\bar\nu_e$'s are directly detected. This is a direct consequence of
Eq.~(\ref{eq:F_A}).

However, once we allow the pinching parameters $\beta_{\nu_\alpha}$ to
vary freely, the accuracy in the determination of the flux parameters
is significantly reduced.  This is represented in
Fig.~\ref{fig:degeneracy1} with shaded areas, yellow (light) and red
(dark) corresponding to 2 and 3$\sigma$ CL respectively.  The striking
consequence is the emergence of a {\em continuous parameter
  degeneracy}\,\footnote{
  Note that this degeneracy is quite different in nature from the
  (discrete) degeneracy one encounters in the determination of lepton
  mixing parameters in neutrino oscillation experiments~
  \cite{BurguetCastell:2001ez,Minakata:2001qm,Fogli:1996pv}.  }:
there is a continuum of allowed fit solutions.  For definiteness, in
what follows we make a very conservative assumption, $\langle
E_{\bar\nu_x}\rangle \leq 32$~MeV.
We find that this degeneracy affects $\bar\nu_e$ and $\nu_x$ flux
parameter determination in a different way.  For $\bar\nu_e$ the
sensitivities to $\langle E_{\bar\nu_e} \rangle$ and $ E^{\rm
  tot}_{\bar\nu_e}$ are reduced to 4\% and 50\%, respectively, at
3$\sigma$ CL.
As can be seen in Fig.~\ref{fig:degeneracy1} for the case of $\nu_x$,
however, the effect is much more drastic.  The region consisting of
degenerate solutions forms a quasi one-dimensional strip extending
mainly in the direction of $\nu_x$ SN flux parameters.  This is a
direct consequence of the different weights with which the original
$\nu_\alpha$ fluxes enter in the observed $\bar\nu_e$ flux: 70\% from
$\bar\nu_e$ and 30\% from $\nu_x$, see Eq.~(\ref{eq:F_A}).


\begin{figure}[h]
\vglue -0.3cm
  \begin{center}
    \includegraphics[width=0.30\textwidth]{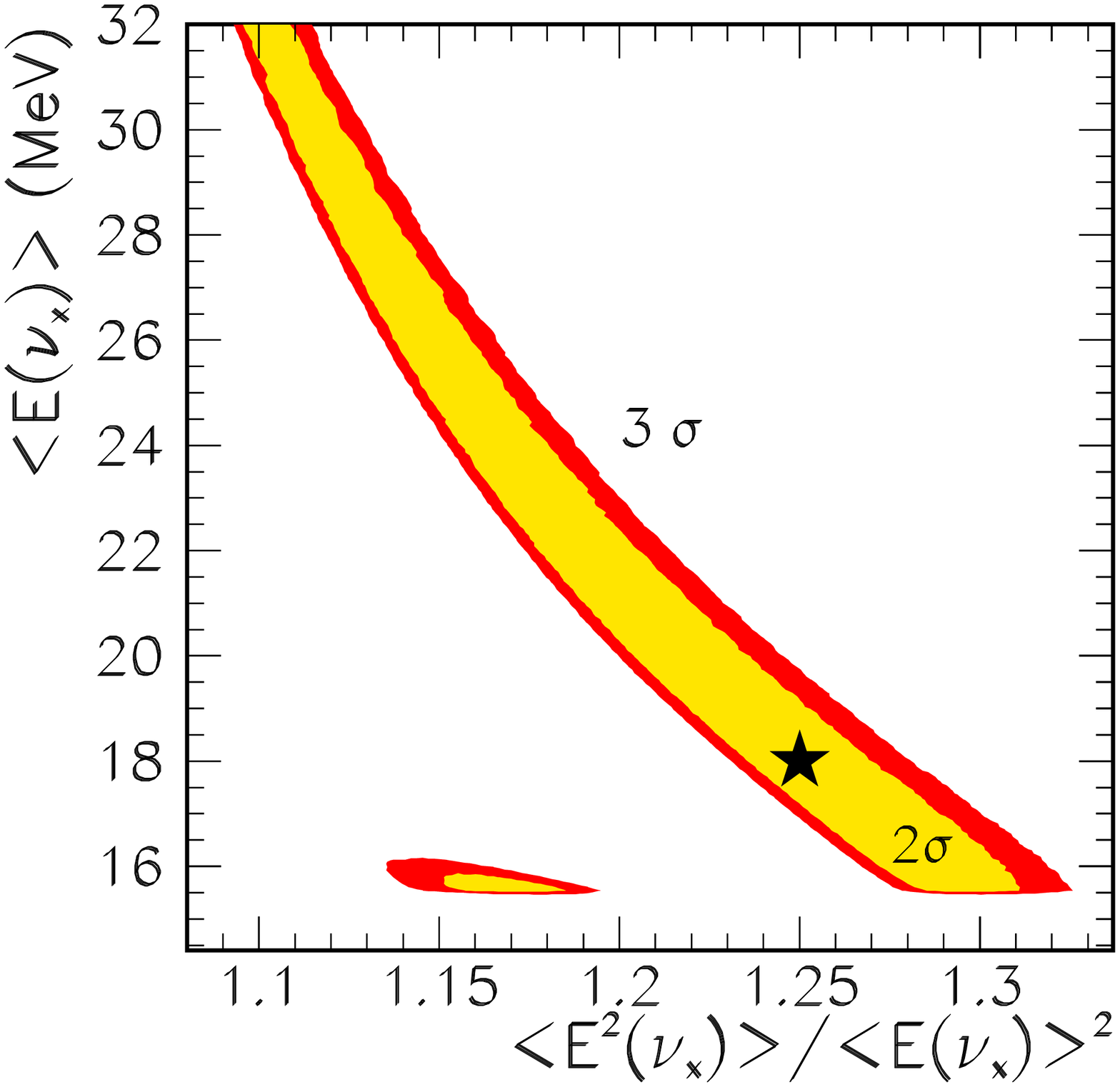}
    \includegraphics[width=0.30\textwidth]{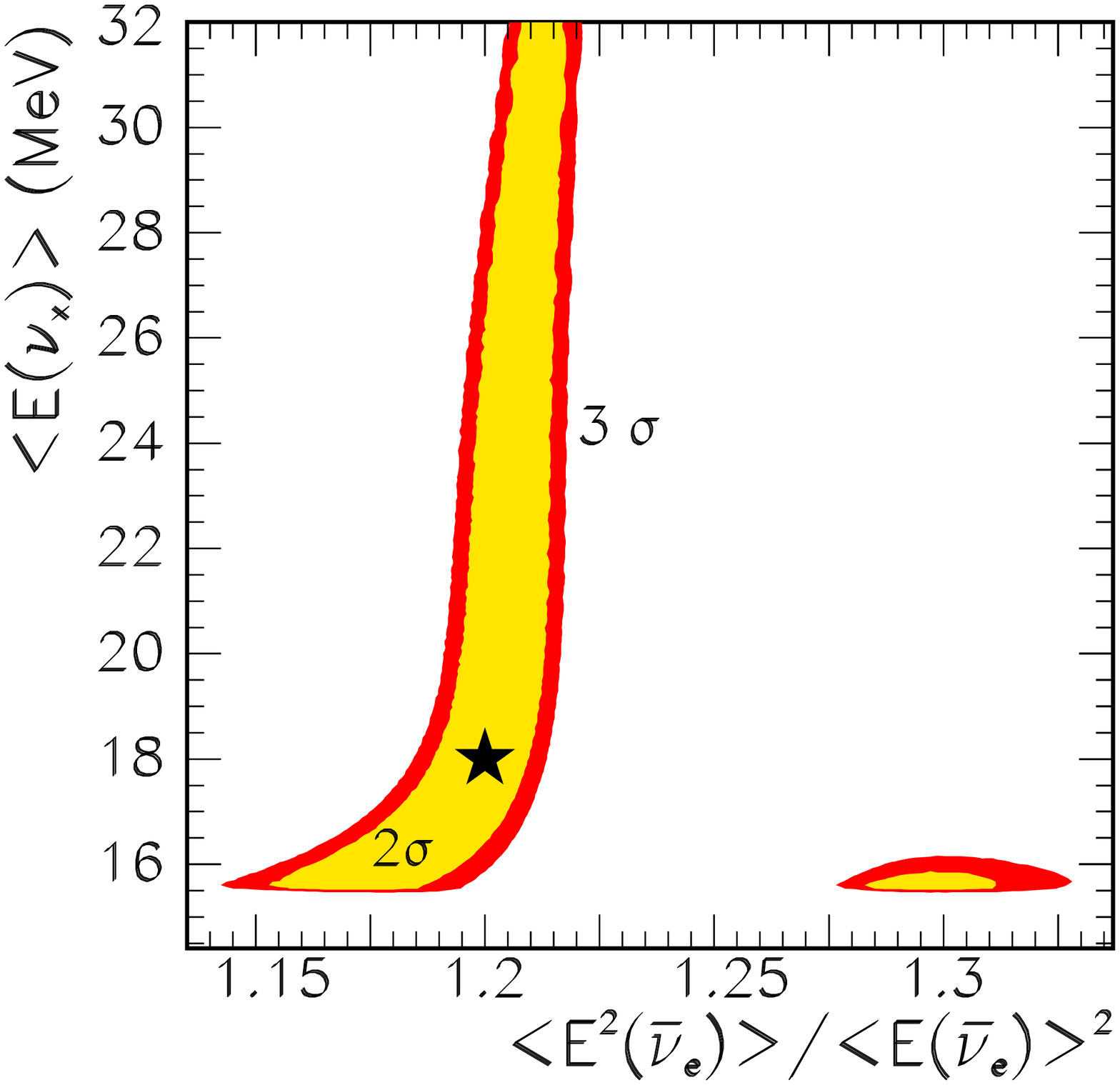}
    \includegraphics[width=0.30\textwidth]{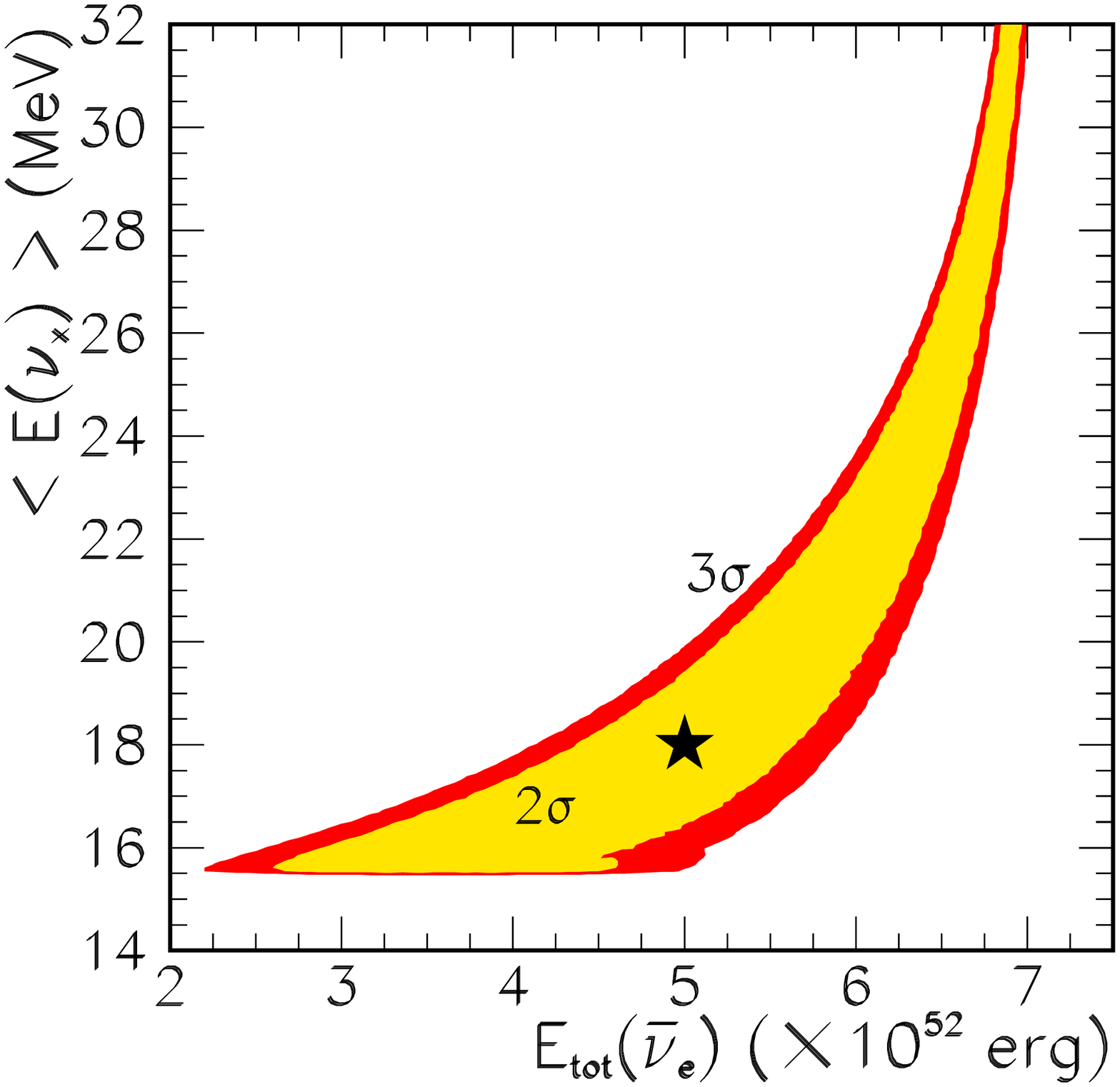}
\vglue -0.1cm
    \includegraphics[width=0.30\textwidth]{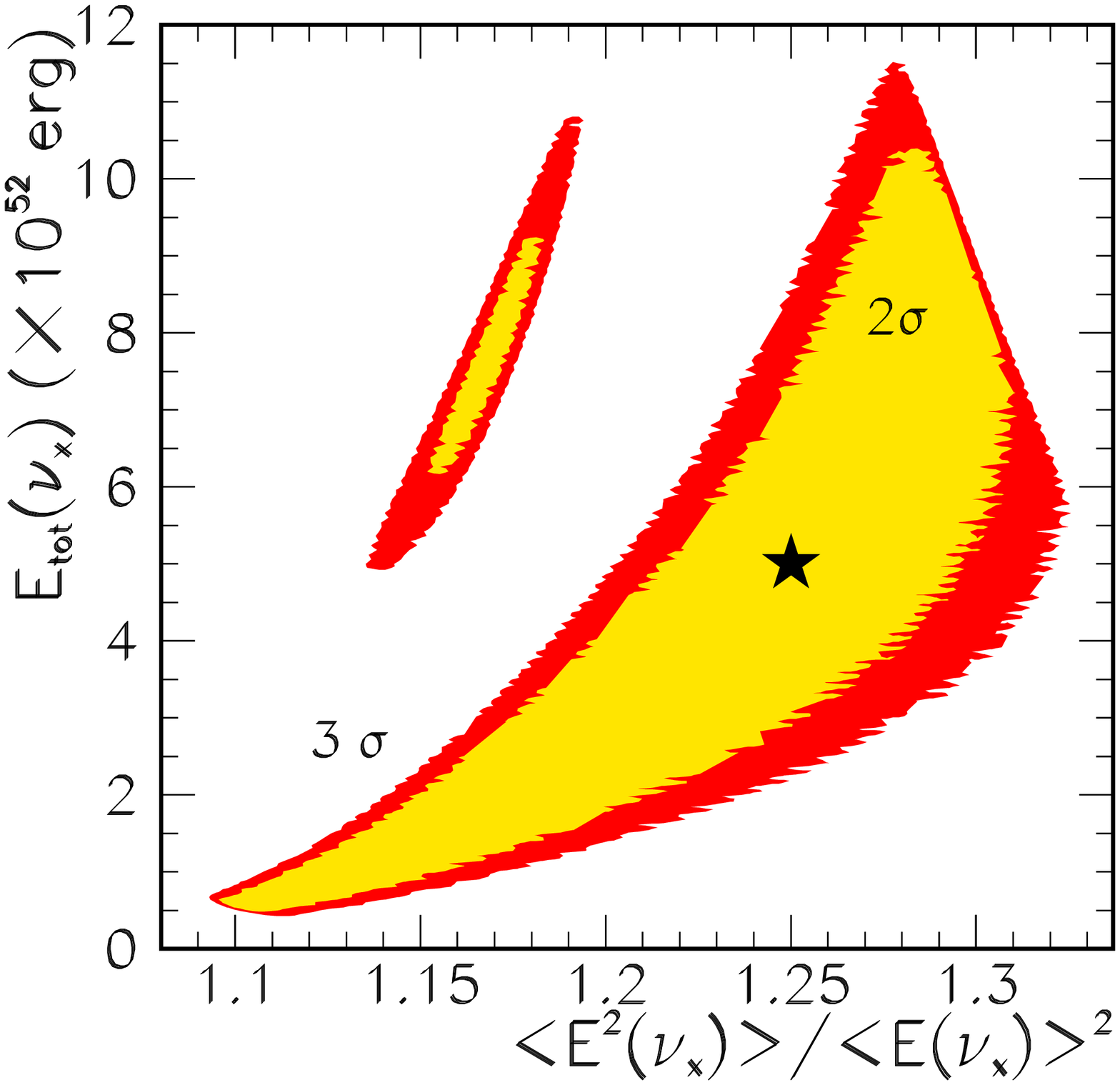}
    \includegraphics[width=0.30\textwidth]{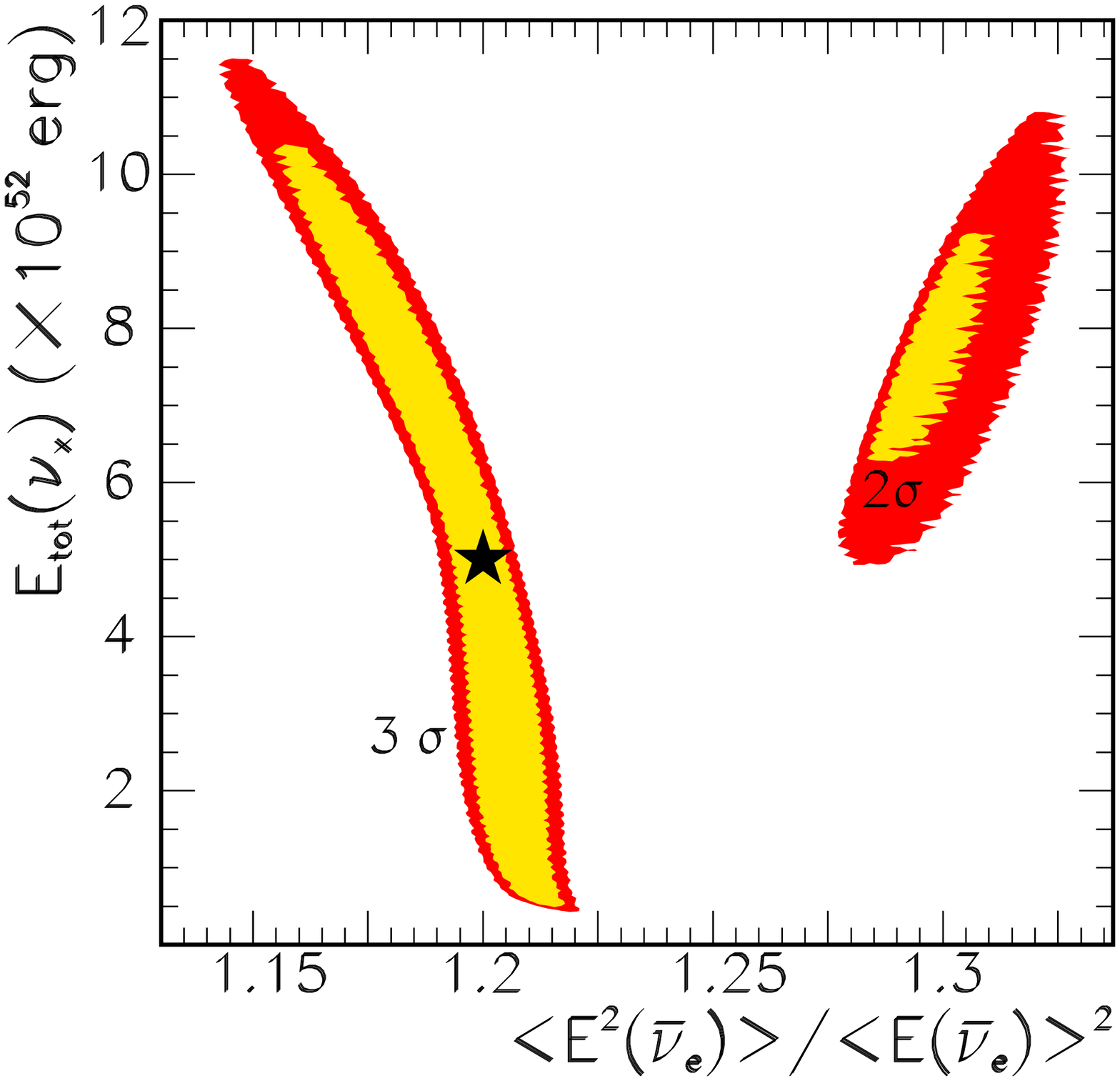}
    \includegraphics[width=0.30\textwidth]{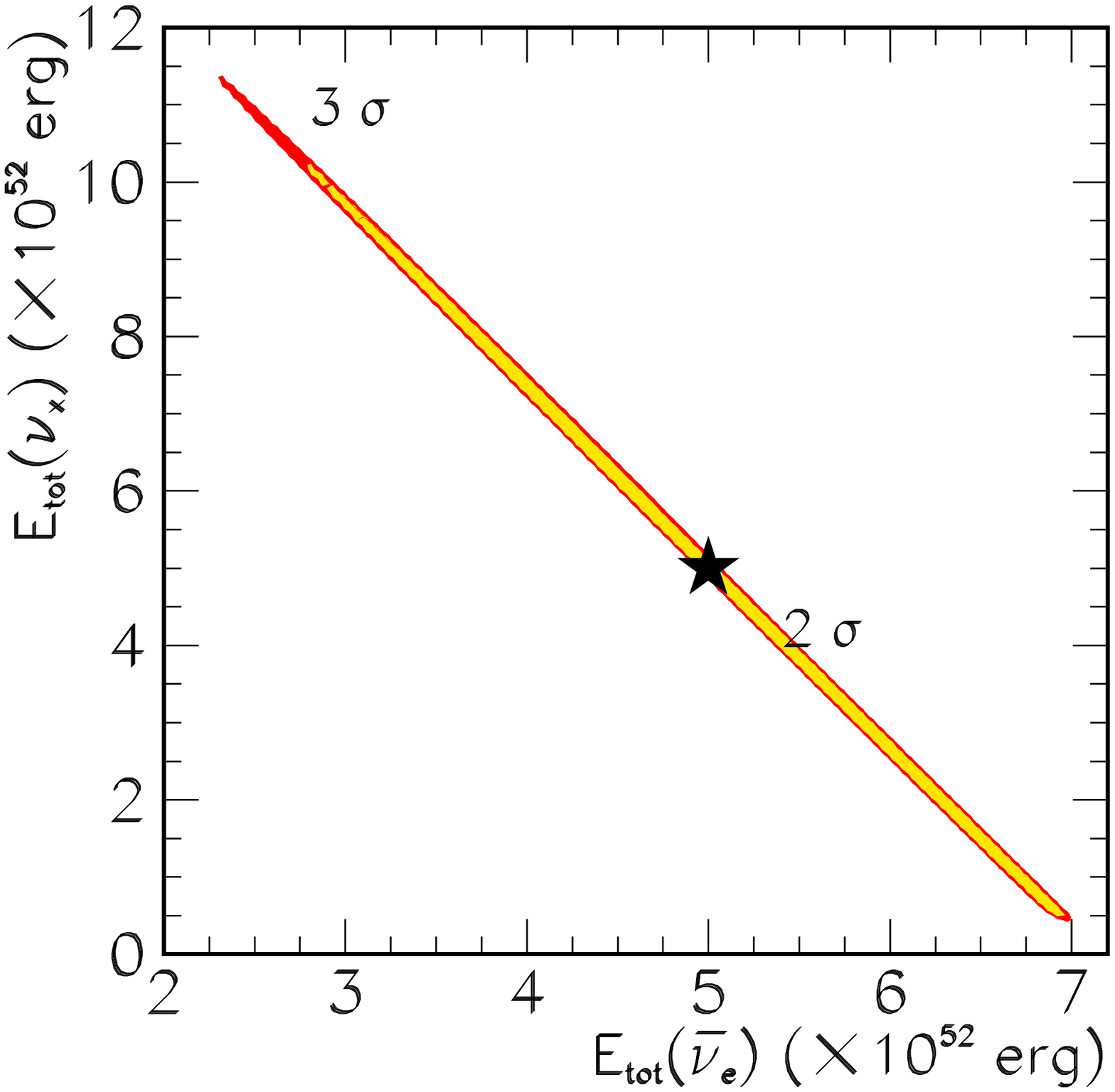}
\vglue -0.1cm
    \includegraphics[width=0.30\textwidth]{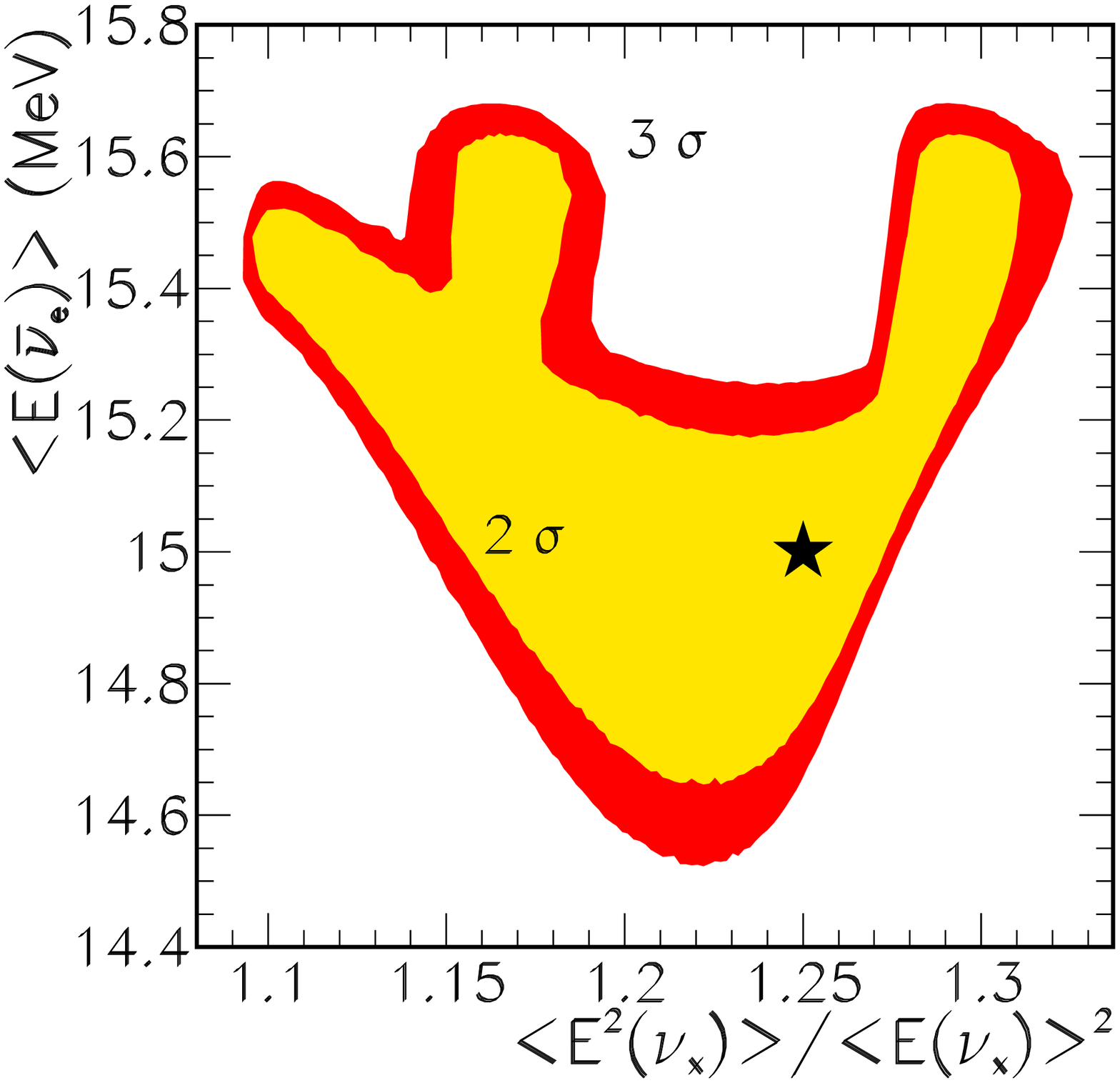}
    \includegraphics[width=0.30\textwidth]{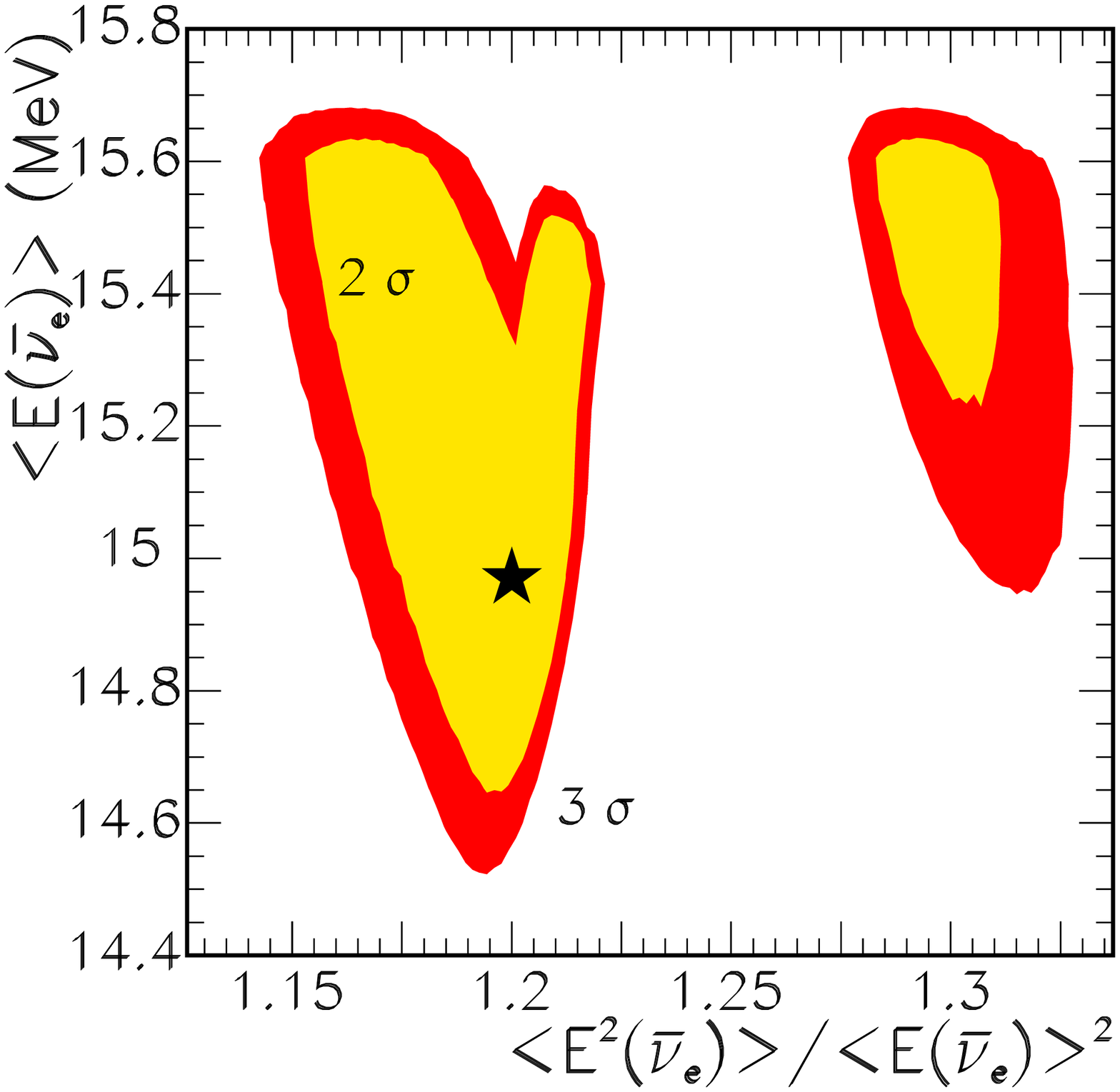}
    \includegraphics[width=0.30\textwidth]{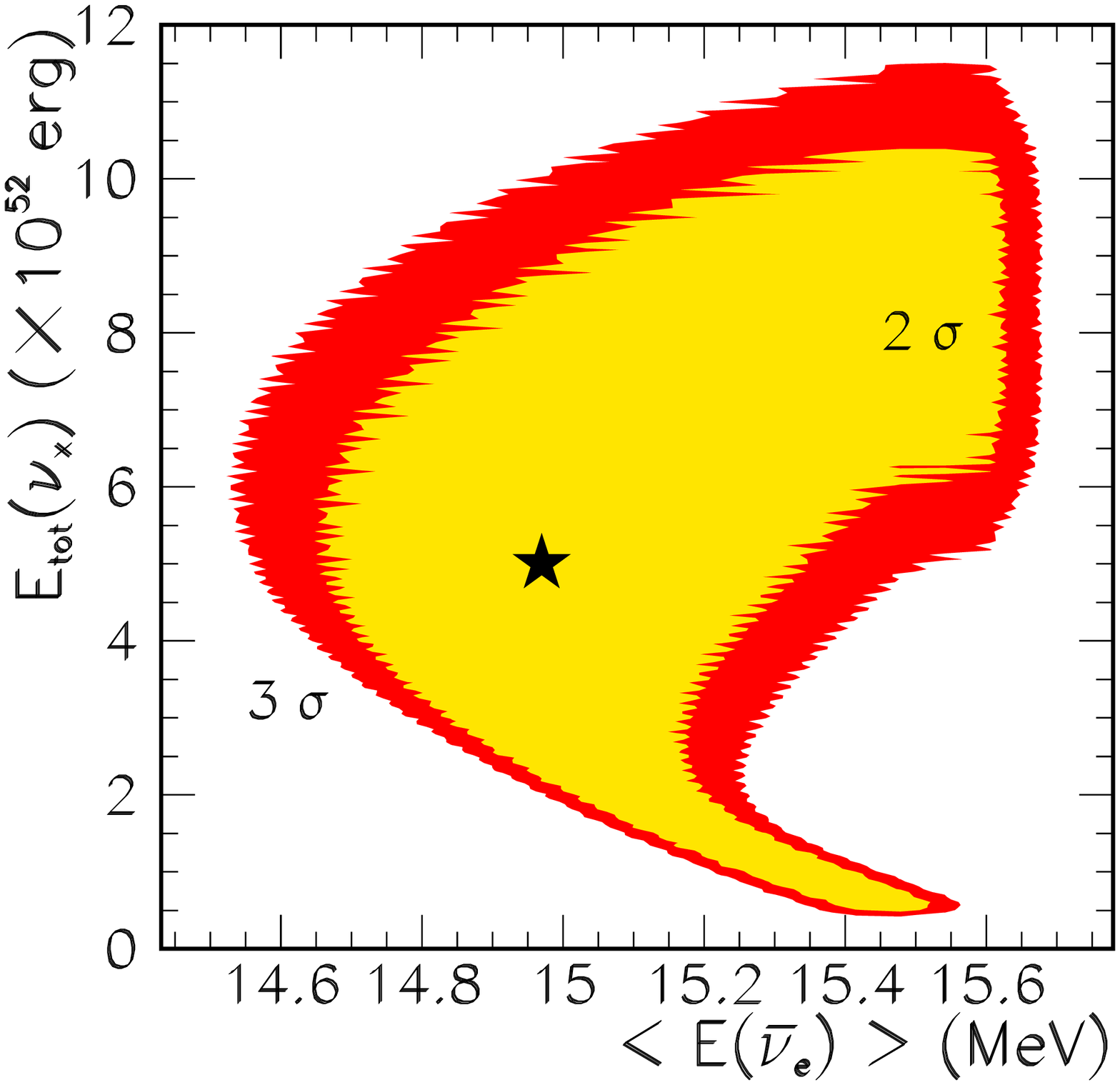}
\vglue -0.1cm
    \includegraphics[width=0.30\textwidth]{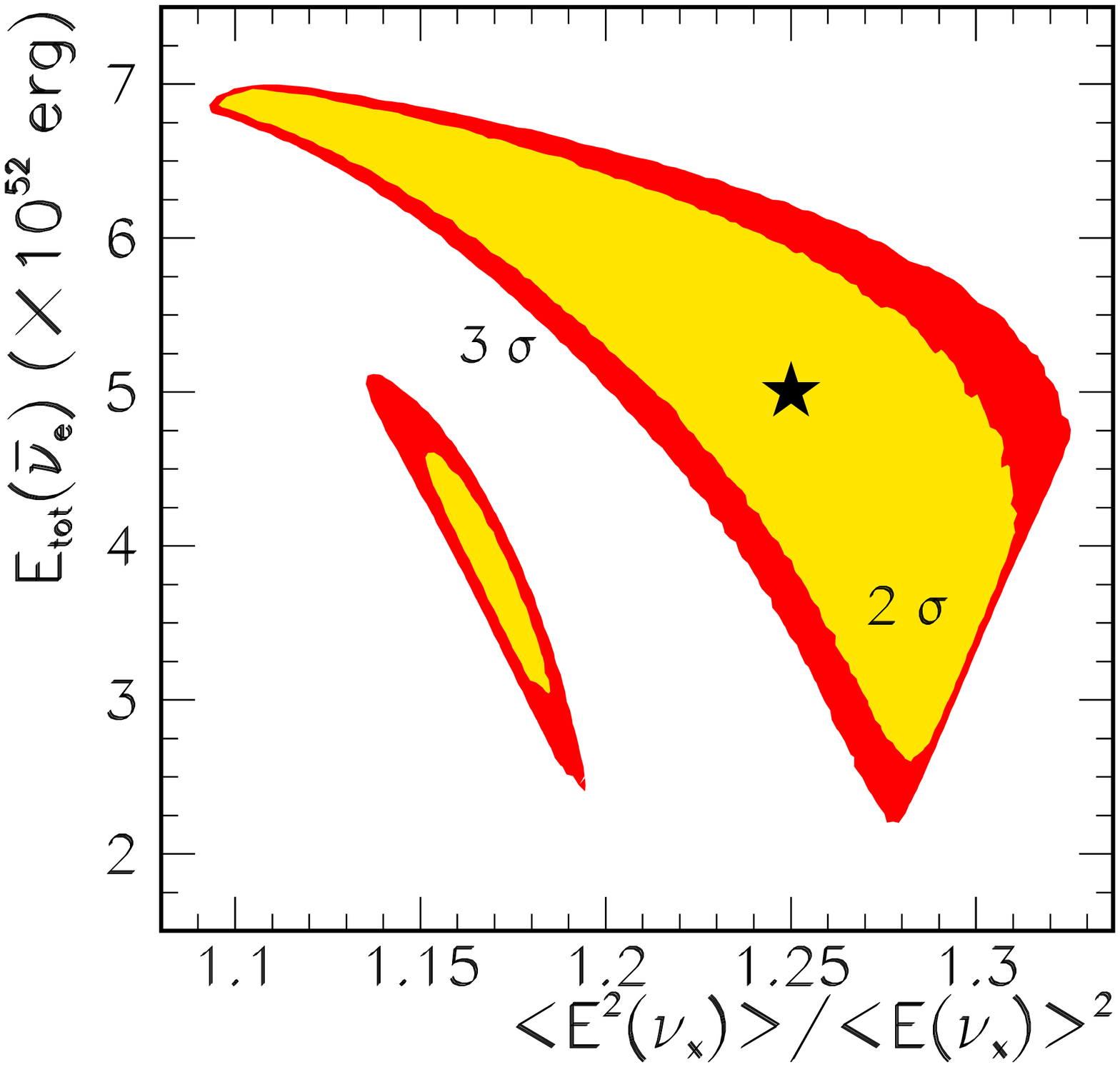}
    \includegraphics[width=0.30\textwidth]{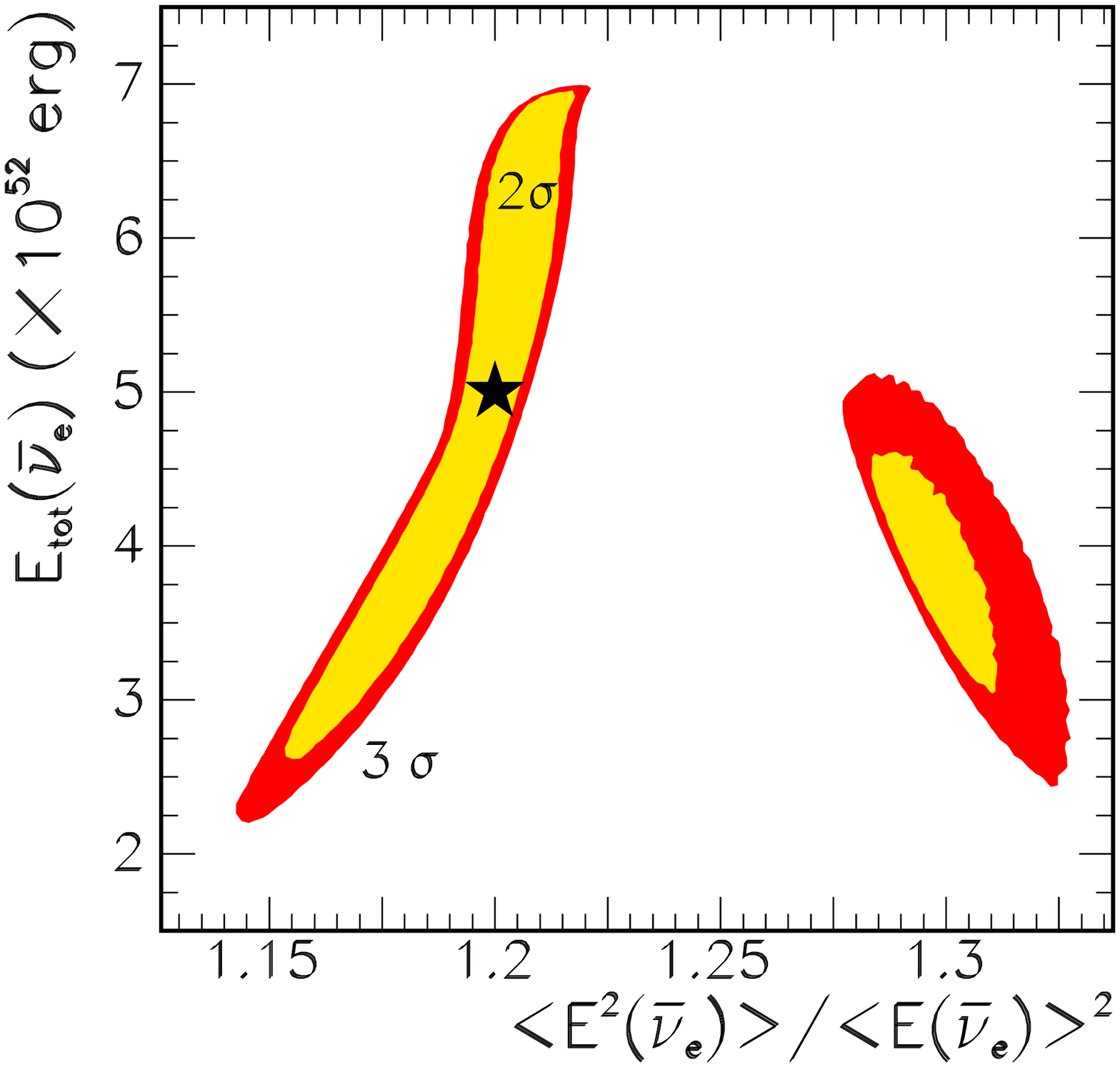}
    \includegraphics[width=0.30\textwidth]{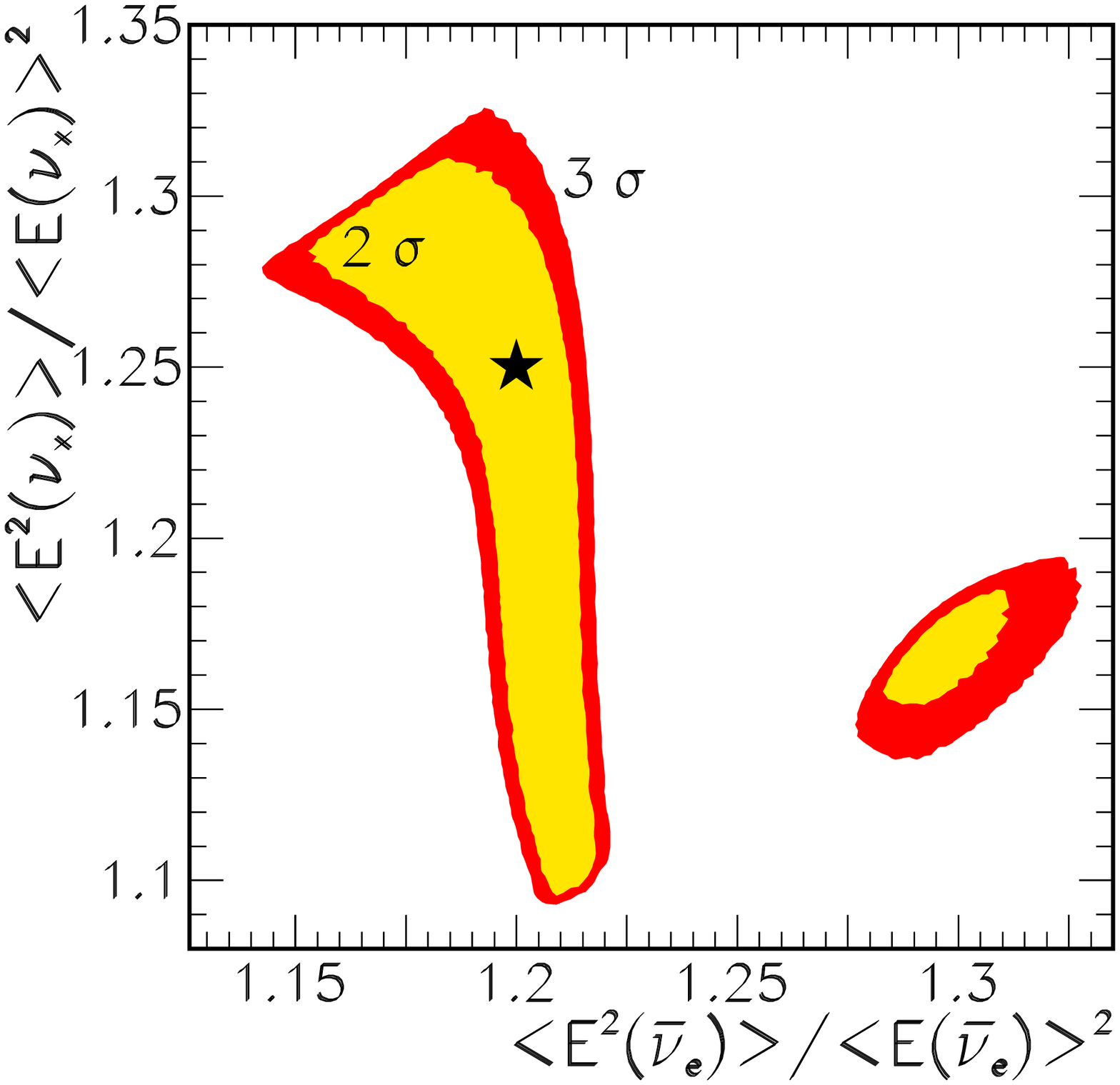}
  \end{center}
\vglue -1.0cm
\caption{ Best-fit point (star) and  2$\sigma$ CL and 3$\sigma$
  CL projections over the displayed parameters for free pinching
  parameters in the Garching parametrization.}
\label{fig:Keil-Keil}
\vglue -0.2cm
\end{figure}


In order to shed more light on the nature of the degeneracy we present
in Fig.~\ref{fig:Keil-Keil} the allowed regions displayed in terms of
various other combinations of the flux parameters. 
>From the last panel of Fig.~\ref{fig:degeneracy1} and the top left
panel of Fig.~\ref{fig:Keil-Keil} it can be inferred that the key of
the continuous degeneracy lies on the possibility of increasing the
pinching of $F^0_{\nu_x}$, i.e. larger $\beta_{\nu_x}$ and smaller
$p_{\nu_x}$. By requiring a stronger pinching of the $\nu_x$ flux,
together with a larger value of $\langle E_{\nu_x}\rangle$ and a
simultaneous reduction of $E^{\rm tot}_{\nu_x}$ it is perfectly
possible to mimic the behavior of observables defined with input
values of $F^0_{\nu_x}$.
In the bottom right panel one can see how the required  variation of
$p_{\nu_x}$ does not necessarily imply a significant modification of
the shape of the $\bar\nu_e$ flux.
This feature can be seen for instance in the left panel of
Fig.~\ref{fig:example}. In that example the {\em fake} spectra can be
made surprisingly similar to the  {\em true} one roughly by keeping
the same original $F_{\bar\nu_e}^0$ parameters and changing
$F_{\nu_x}^0$ following the previous recipe: a reduction of $E^{\rm tot}_{\nu_x}$
together with an increase of $\beta_{\nu_x}$ and $\langle E_{\nu_x}\rangle$.

Another salient feature that can be observed in most of the panels of
Fig.~\ref{fig:Keil-Keil} is an island structure, i.e. the presence
of a region separated from the main allowed region
containing the best fit point. To understand the origin of such
structure one has to realize that this region arises for larger values
$p_{\bar\nu_e}$ and $\langle E_{\bar\nu_e}\rangle$ than the initial
ones, and smaller $p_{\nu_x}$ and $\langle E_{\nu_x}\rangle$.  This
is related to the trivial degeneracy mentioned before, the extra
solution obtained by interchanging the $\bar\nu_e$ and $\nu_x$
spectra.
When we do not impose the condition $\langle E_{\nu_x} \rangle >
\langle E_{ \bar\nu_e} \rangle$, the allowed region has a ``cross
shape'' because of the trivial degeneracy, namely the region with
swapped parameters between $\bar\nu_e$ and $\nu_x$ is allowed.  It
appears to us that the island which is left over
is a remnant of the swapped parameter region. 

\subsection{Robustness of the degeneracy}  
\label{subsec:robustness}

\begin{figure}[h]
  \begin{center}
    \includegraphics[width=0.32\textwidth]{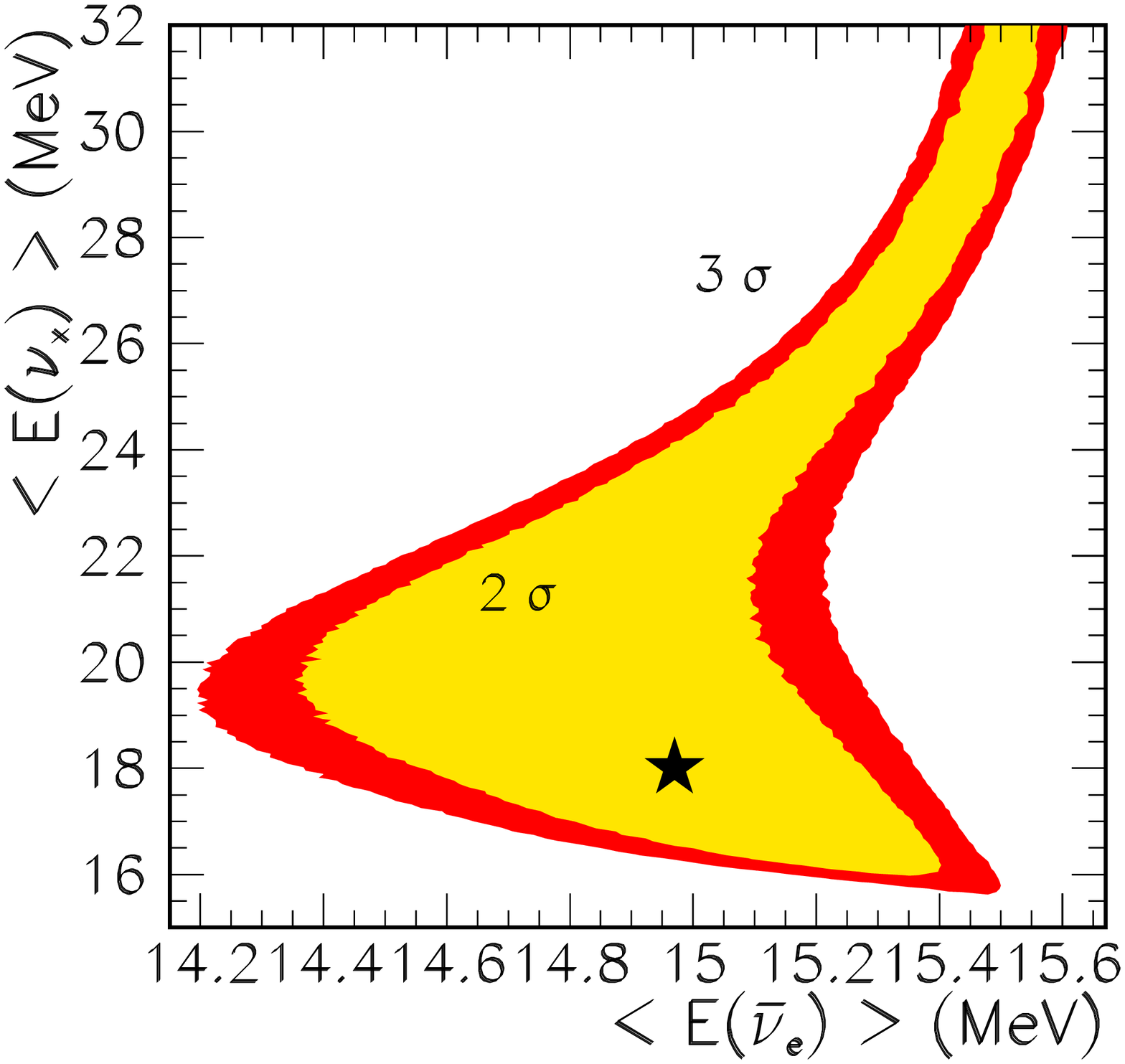}
    \includegraphics[width=0.32\textwidth]{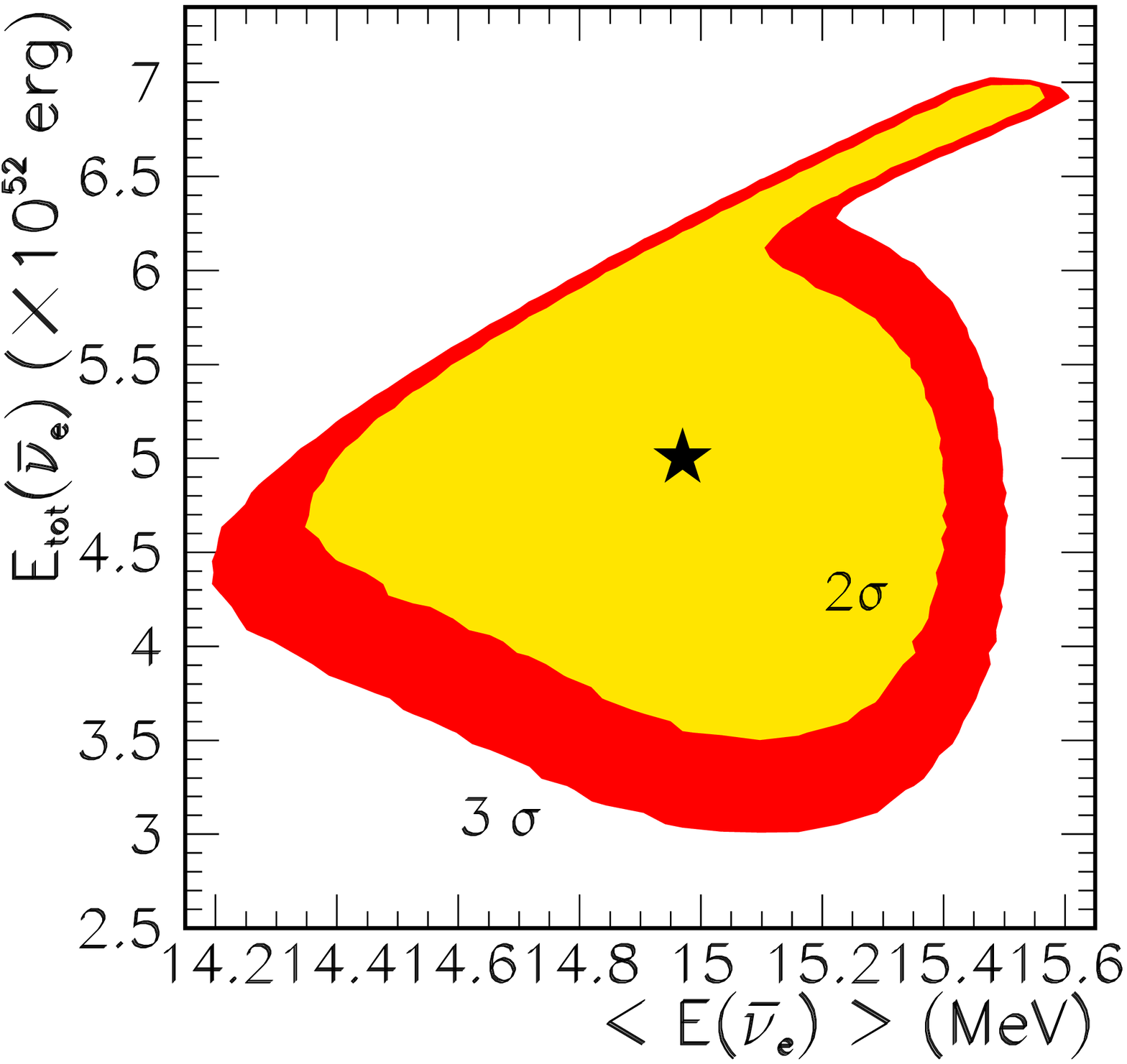}
    \includegraphics[width=0.32\textwidth]{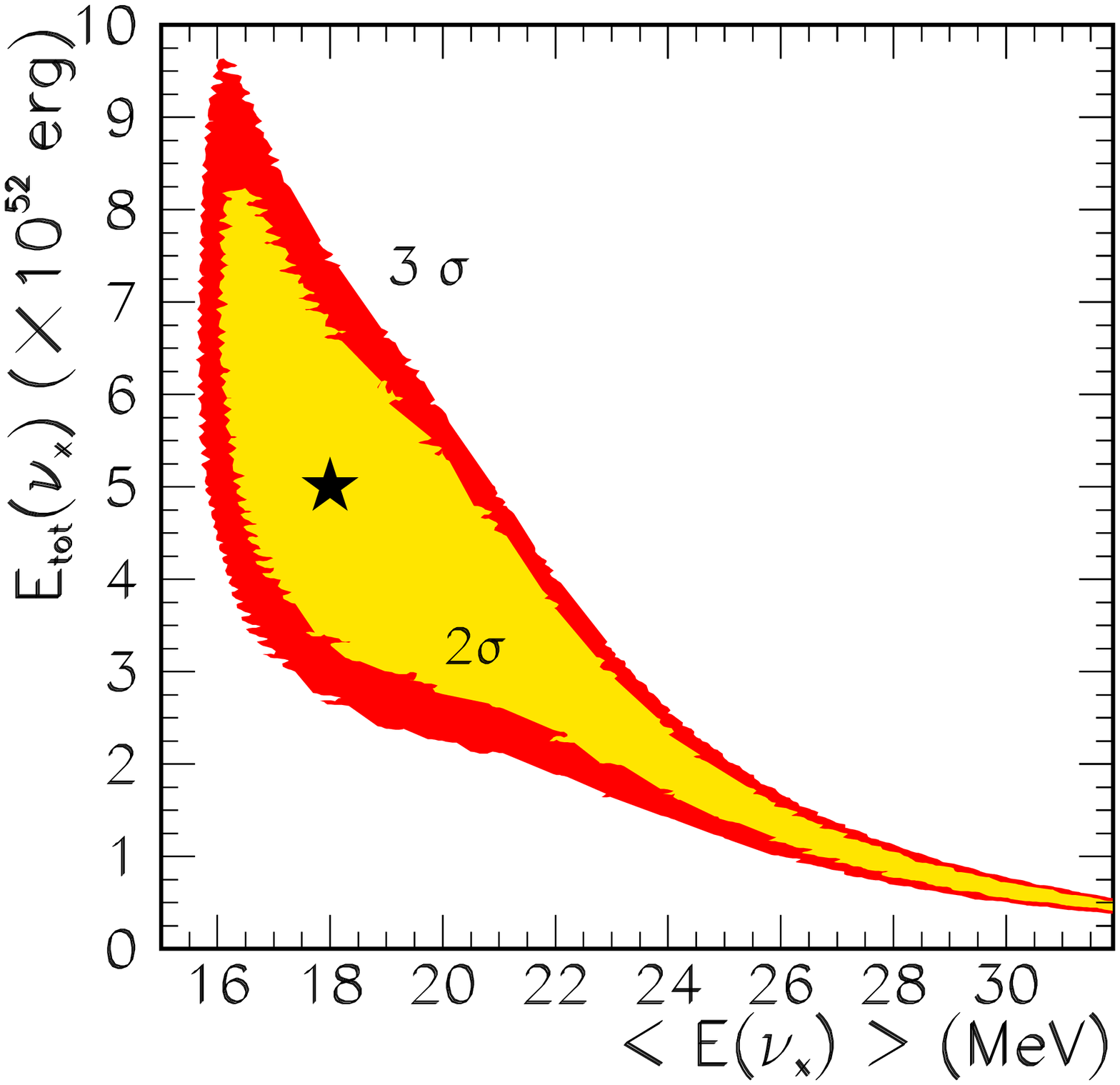}
    \includegraphics[width=0.32\textwidth]{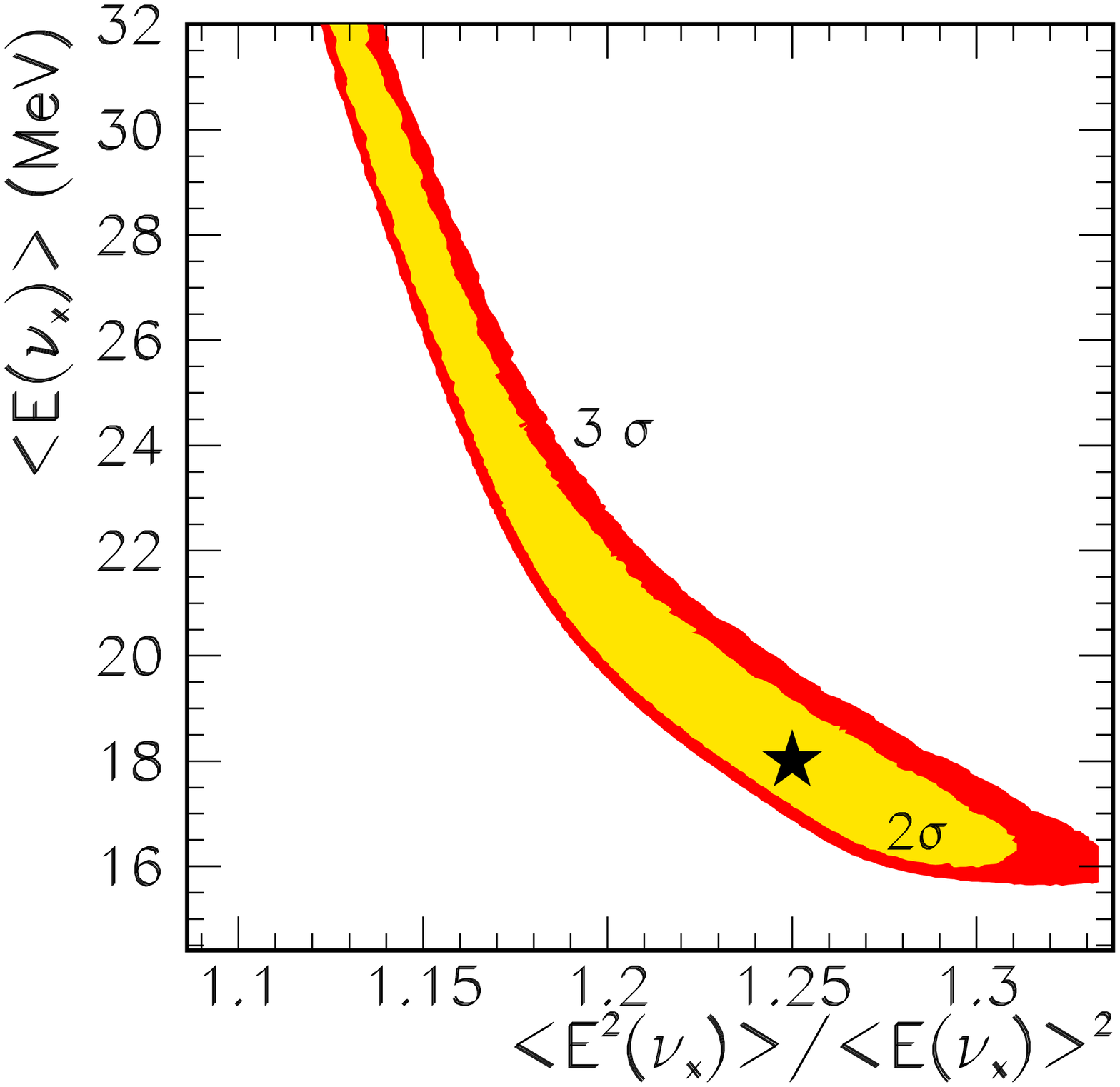}
    \includegraphics[width=0.32\textwidth]{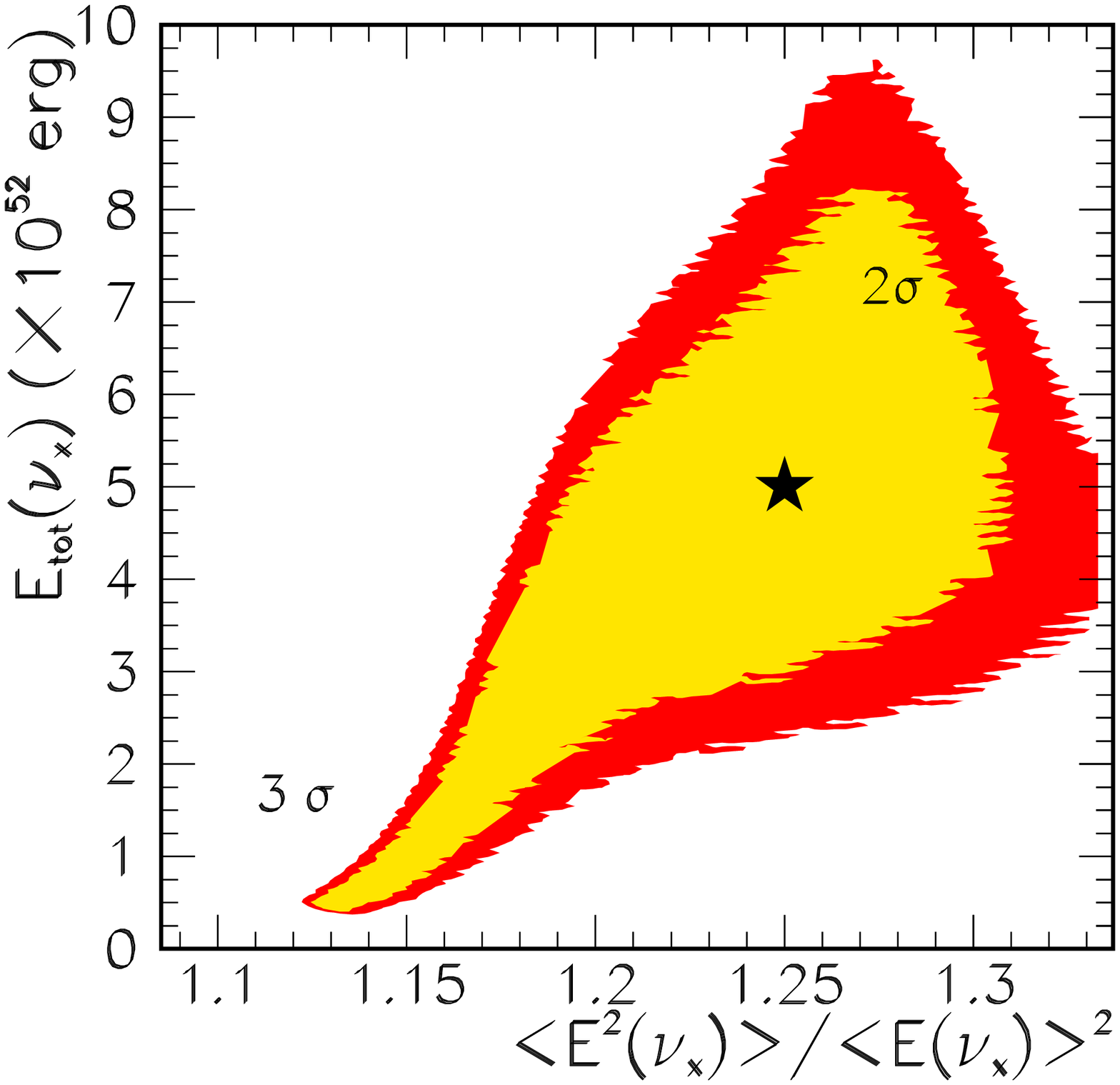}
    \includegraphics[width=0.32\textwidth]{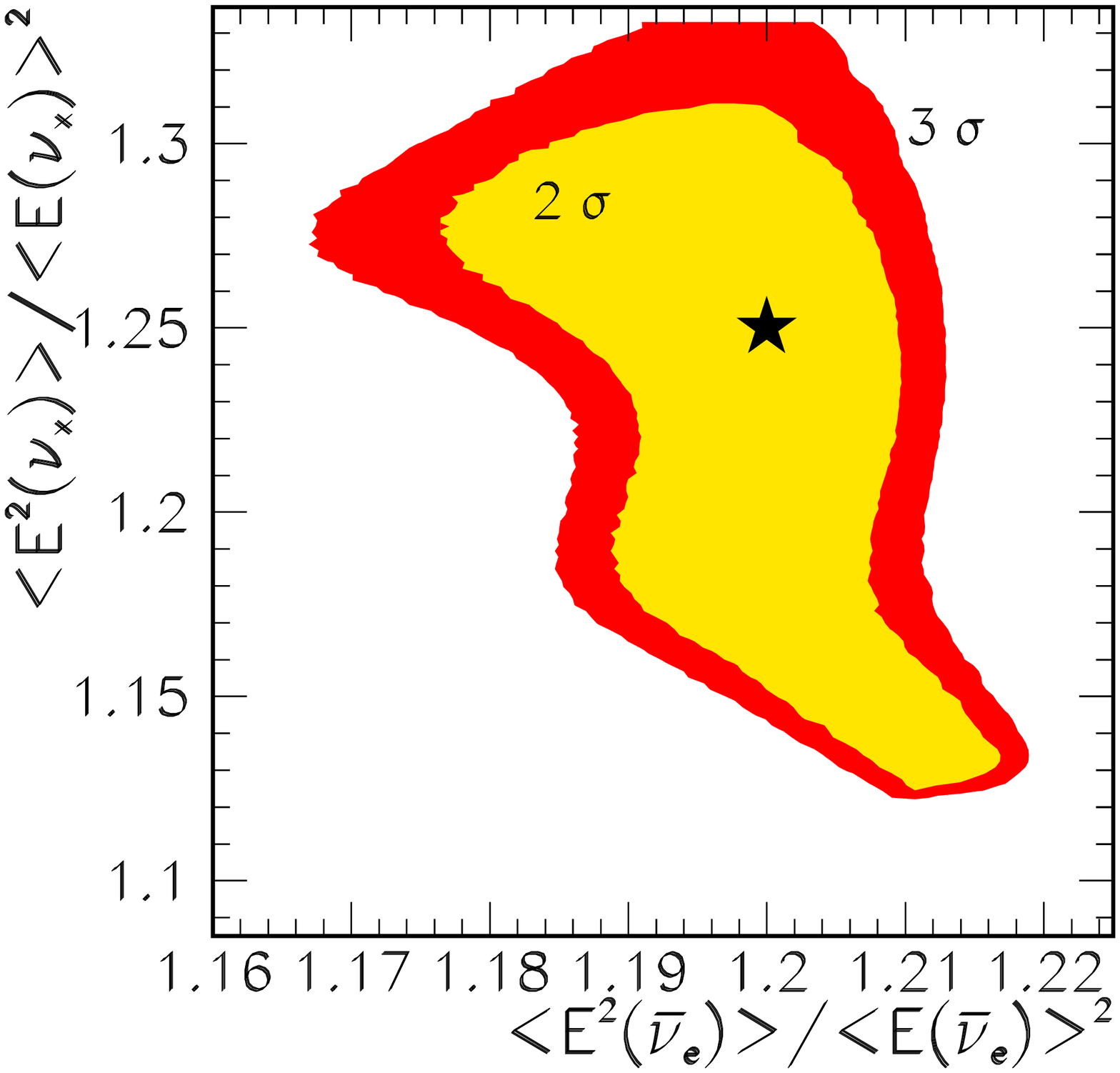}
  \end{center}
\vglue -0.9cm
\caption{Best-fit points (star) and 2$\sigma$ CL and 3$\sigma$ CL
  projections over the displayed parameters for free pinching
  parameters in the Fermi-Dirac parametrization. Here six
  representative combinations of fitting parameters were chosen,
  corresponding to the 3 panels in Fig.~\ref{fig:degeneracy1} and top
  3 panels in Fig.~\ref{fig:Keil-Keil}.  .}
\label{fig:FD-FD}
\end{figure}

It is reasonable to ask whether the degeneracy is an artifact of the
particular parametrization we employ, or an accidental consequence of
the particular choice of the initial parameters.
In order to answer the first question we performed the same exercise
by using the pinched Fermi-Dirac parametrization,
Eq. (\ref{eq:flux-FD}), both in the preparation of data, using the
input values given in Tab.~\ref{tab:input}, as well as in fitting
them.
As shown in Fig.~\ref{fig:FD-FD} the shapes and the sizes of the
allowed regions are very similar to those given in
Figs.~\ref{fig:degeneracy1} and~\ref{fig:Keil-Keil} (for brevity, we
have shown our results only for 6 representative combinations out of
15, because the others are similar to previous plots).  The continuous
parameter degeneracy is also present for the same combination of
$\nu_x$ parameters: strong pinching, large mean energies and small
integrated luminosity.
The main difference from the case with the Garching 
parametrization is the absence of the island structure.


\begin{figure}[h]
  \begin{center}
    \includegraphics[width=0.32\textwidth]{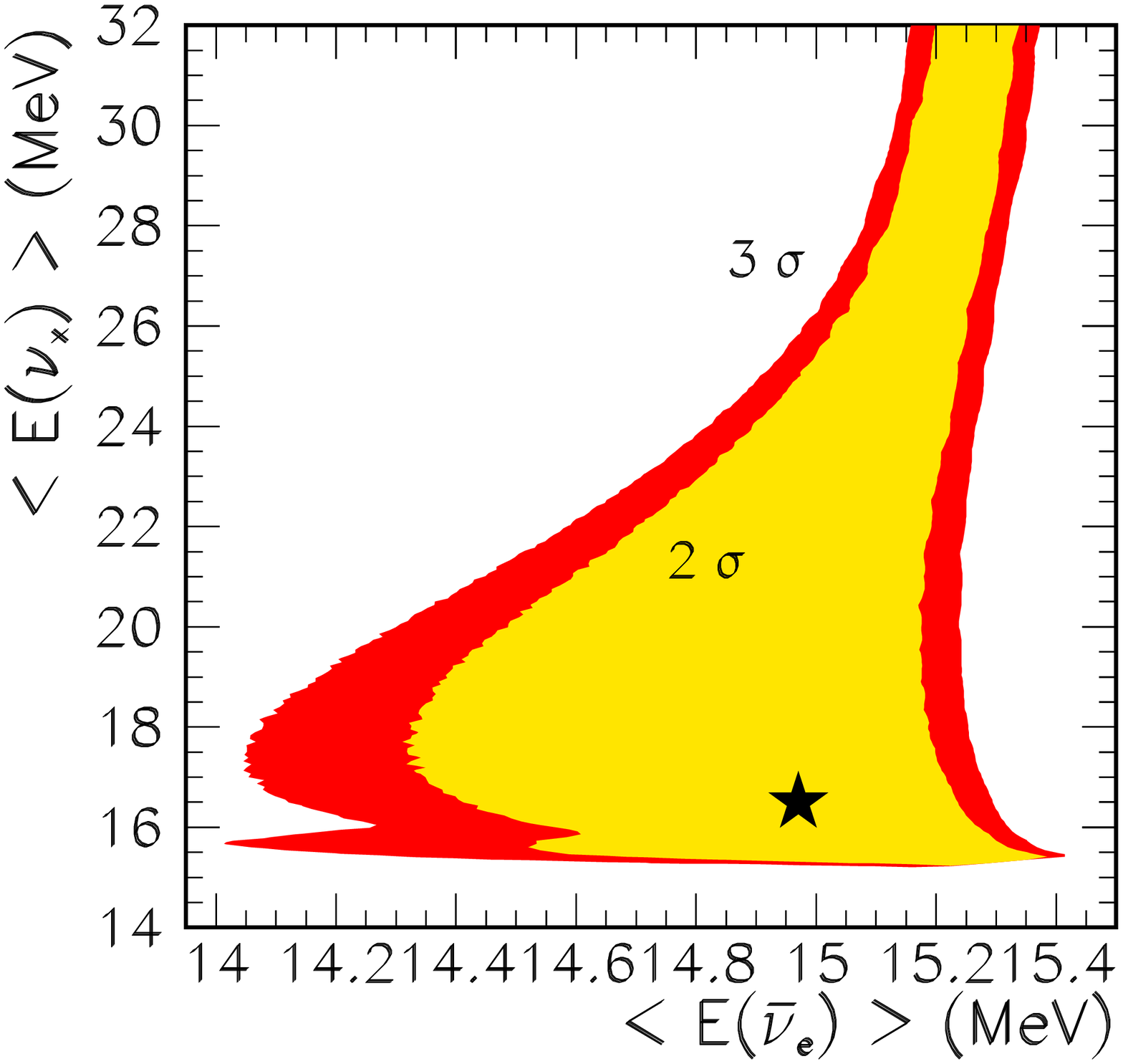}
    \includegraphics[width=0.32\textwidth]{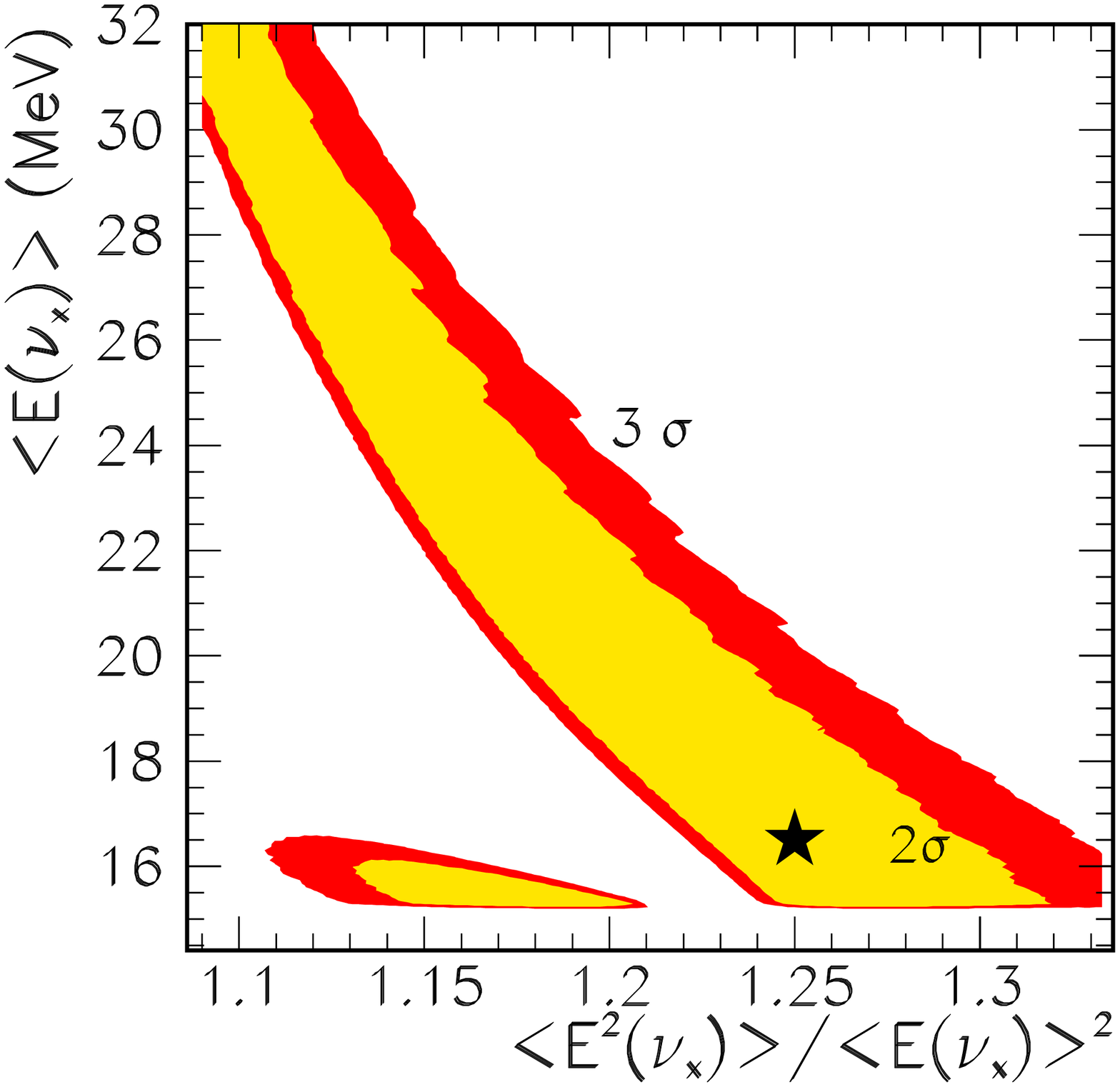}
  \end{center}
\vglue -0.9cm
\caption{Best-fit point (star) and allowed 2$\sigma$ CL and 3$\sigma$
  CL projections over the displayed parameters for free pinching
  parameters in the Garching parametrization. Here we use as input the
  values given in Tab.~\ref{tab:input}, except for $\langle E_{\nu_x}
  \rangle$, which is 16.5 MeV.  Two fitting parameter combinations
  were selected: the left panel corresponds to the top left panel of
  Fig.~\ref{fig:FD-FD} whereas the right one corresponds to the bottom
  left panel of the same figure.}
\label{fig:KK16.5}
\end{figure}

\begin{figure}[h]
  \begin{center}
    \includegraphics[width=0.32\textwidth]{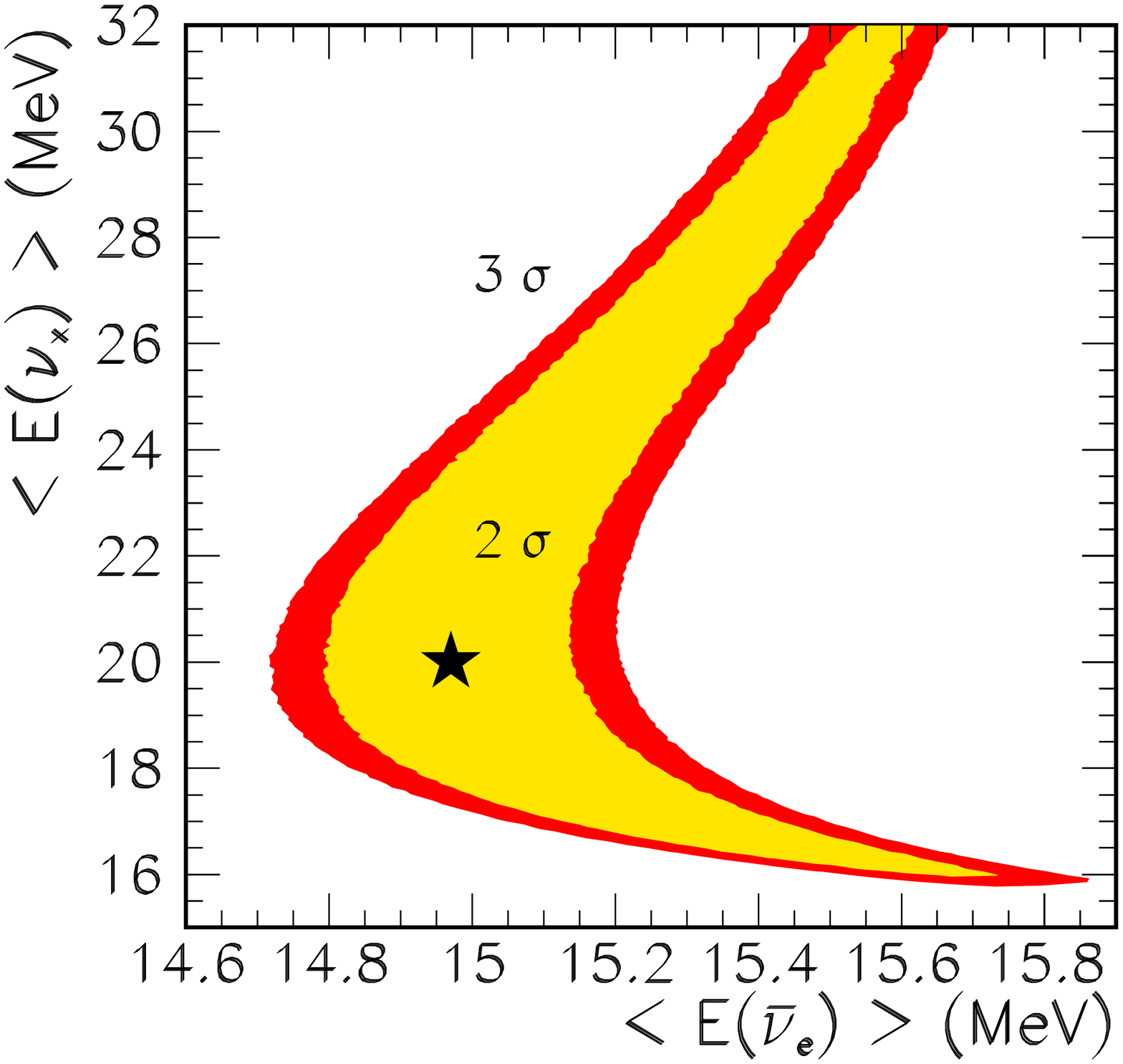}
    \includegraphics[width=0.32\textwidth]{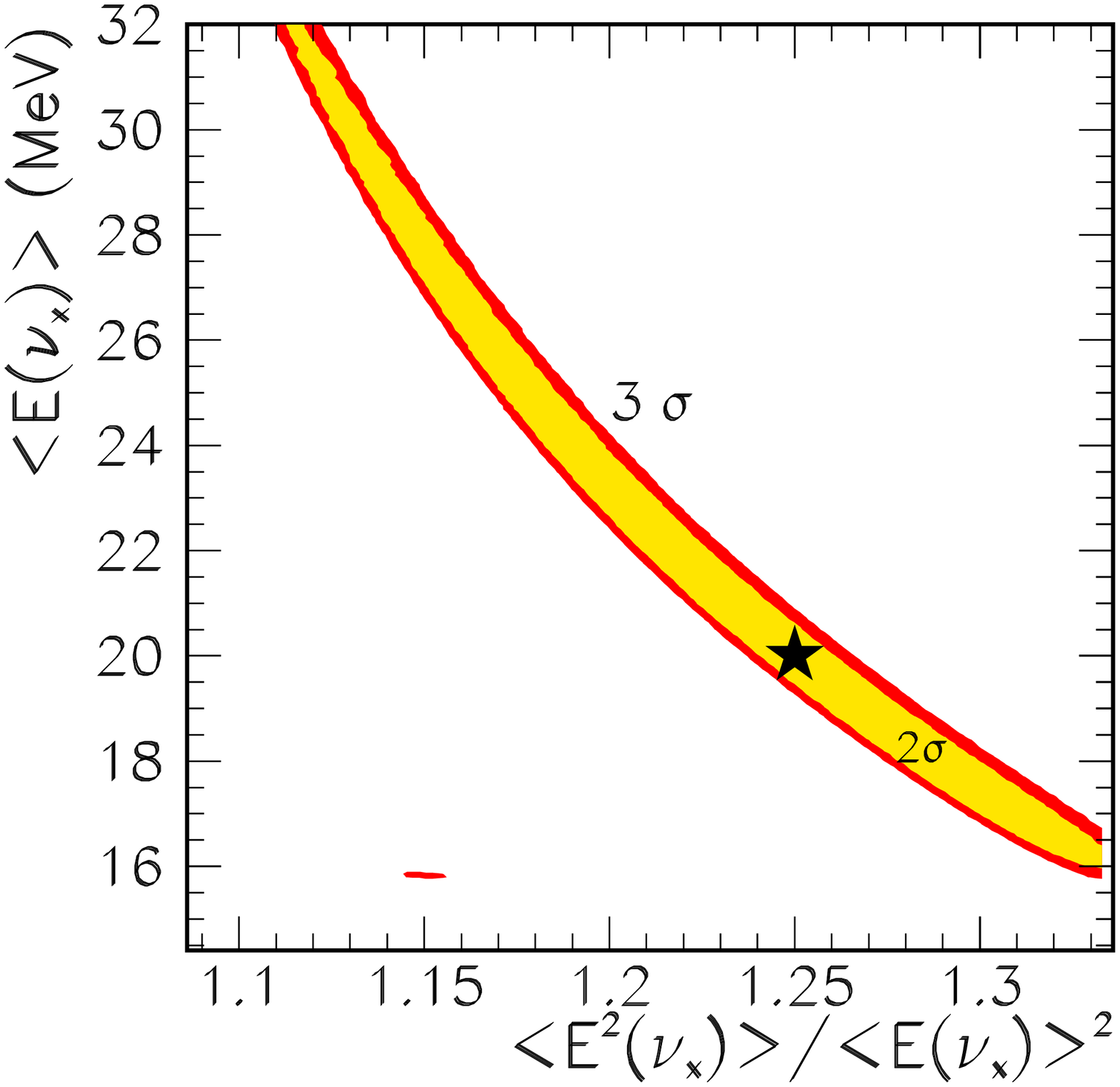}
  \end{center}
\vglue -0.9cm
\caption{Same as in Fig.~\ref{fig:KK16.5} but for $\langle E_{\nu_x}
  \rangle=20$~MeV.}
\label{fig:KK20}
\end{figure}

In order to shed some light on the dependence of the results on the
values of the initial neutrino fluxes we have redone the fit in the
case of Garching data fitted by Garching distribution for $\langle
E_{\nu_x} \rangle=16.5$ and 20~MeV, which are presented in
Figs.~\ref{fig:KK16.5} and~\ref{fig:KK20}, respectively.
One can see the shape of the allowed regions changes but the presence
of the continuous degeneracy persists with a similar range of $\nu_x$
parameters.
On the other hand, the size of the island region changes depending on
the initial parameters considered.  The more similar the initial
$\bar\nu_e$ and $\nu_x$ spectra, the bigger the island.  This is due
to the fact that it is easier to interchange the role of the two
flavors.
%
%

For completeness we have also varied the number of energy bins used in
the statistical analysis, but found no significant change in the
result.
We conclude then that the presence of the continuous degeneracy is a
robust feature of the reconstruction analysis of supernova parameters.

\subsection{Effect of uncertainties in the SN neutrino flux spectra}
\label{cross-fit}

So far we have assumed in our analysis that we know the functional
form of the supernova neutrino flux spectra prepared by the exploding
star.  Of course, this is {\em not} the case.  What would be the
effect of our ignorance of primordial SN neutrino flux spectra
parametrization on the analysis?  In order to gain some insight on
this issue in this subsection we attempt a new procedure: to generate
the data with a Fermi-Dirac distribution and to fit it assuming the
Garching parametrization, and vice versa.
This way we try to check not only whether it is possible to determine
the flux parameters, but also its functional form.
\vglue 1cm

\begin{figure}[h!]
  \begin{center}
    \includegraphics[width=0.33\textwidth,angle=0]{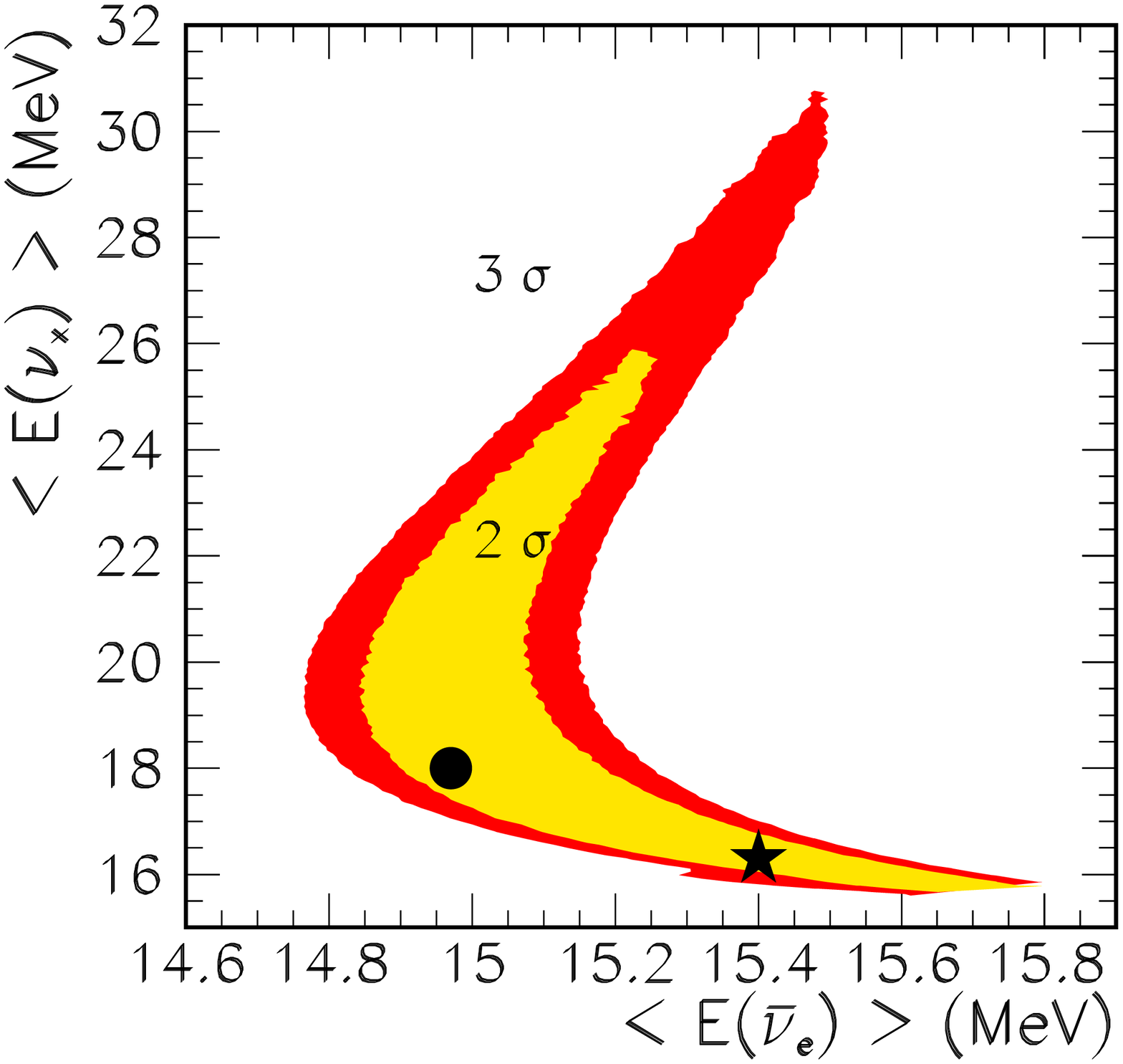}
    \includegraphics[width=0.33\textwidth,angle=0]{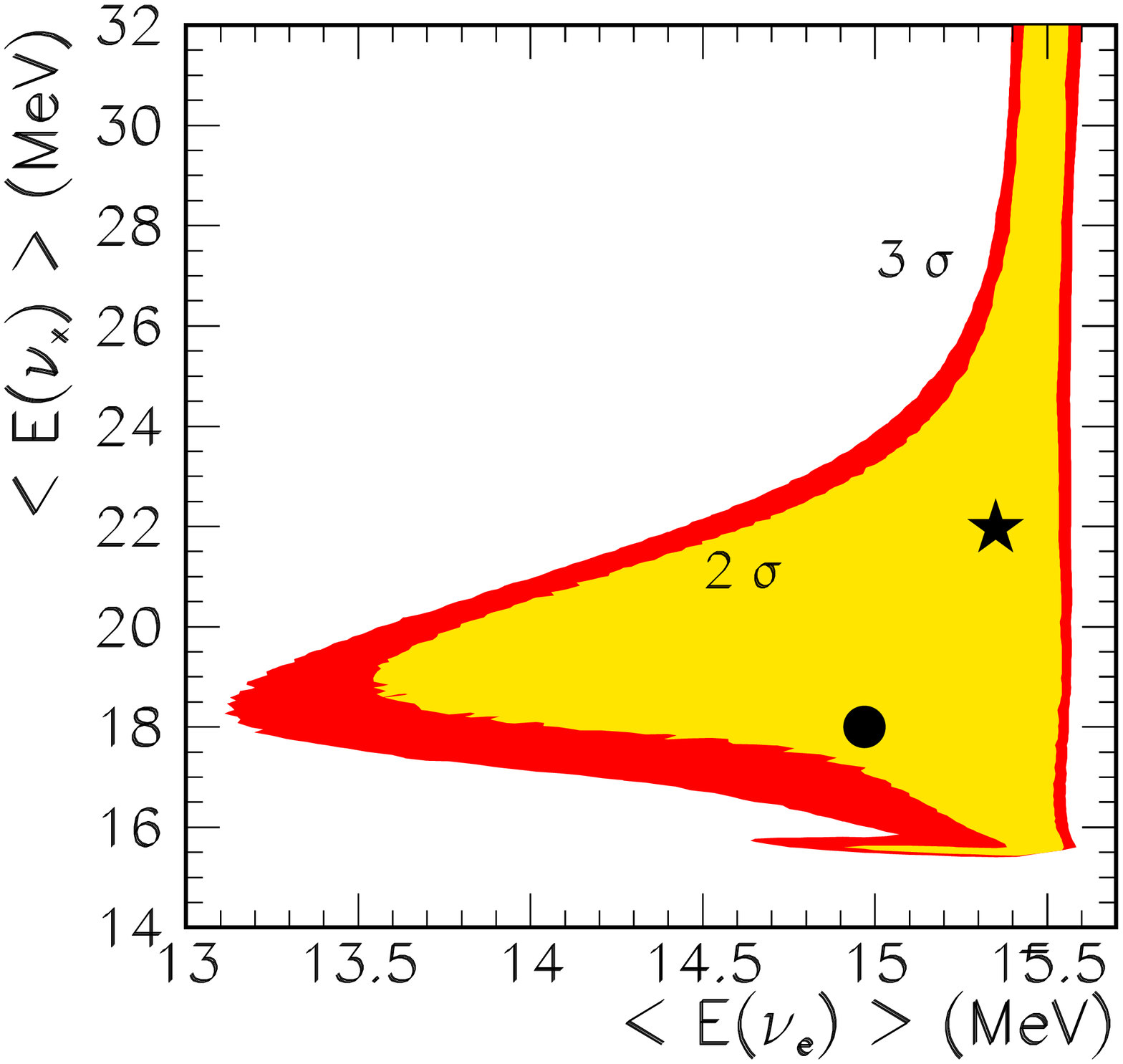}
    \includegraphics[width=0.33\textwidth,angle=0]{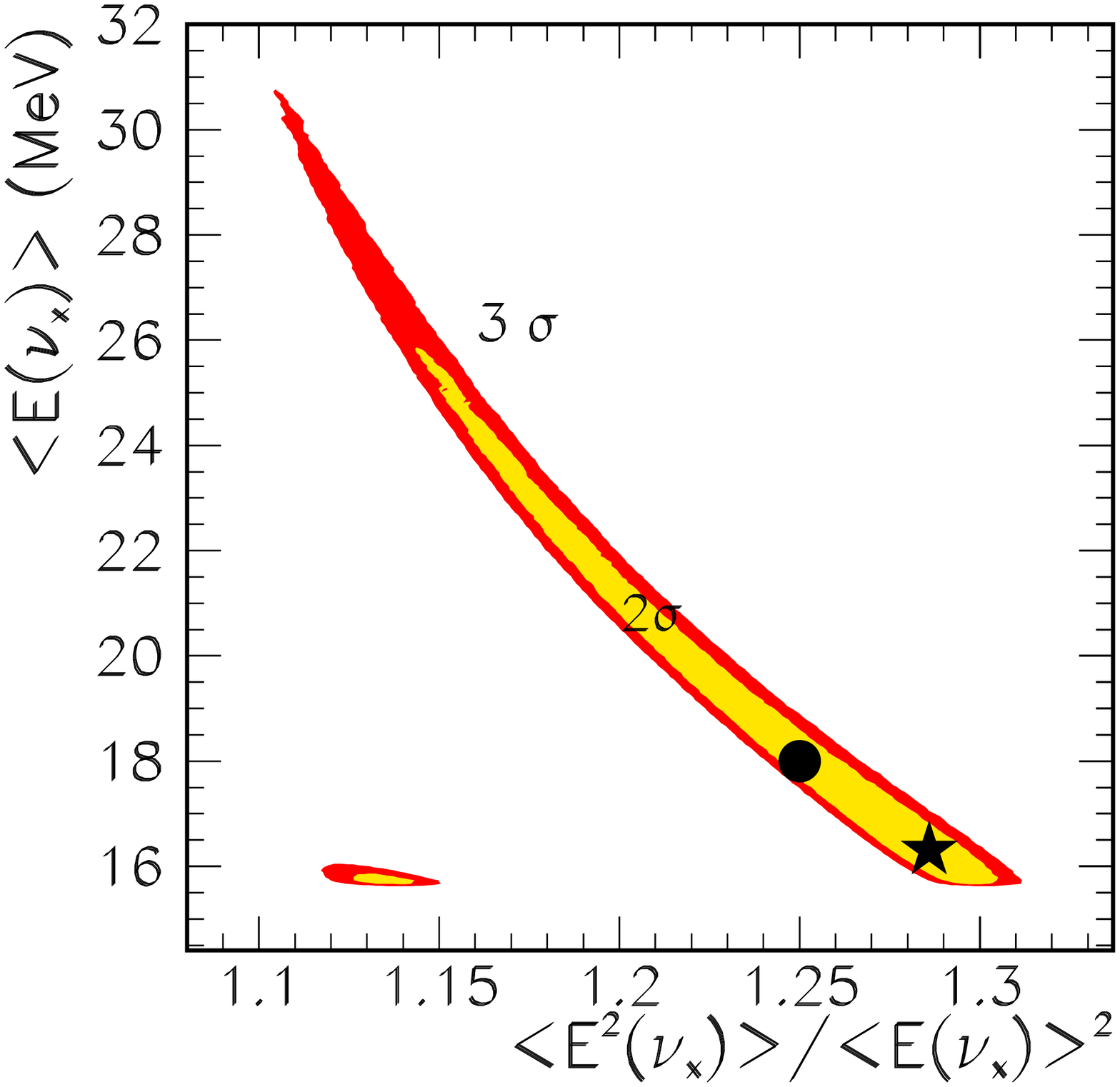}
    \includegraphics[width=0.33\textwidth,angle=0]{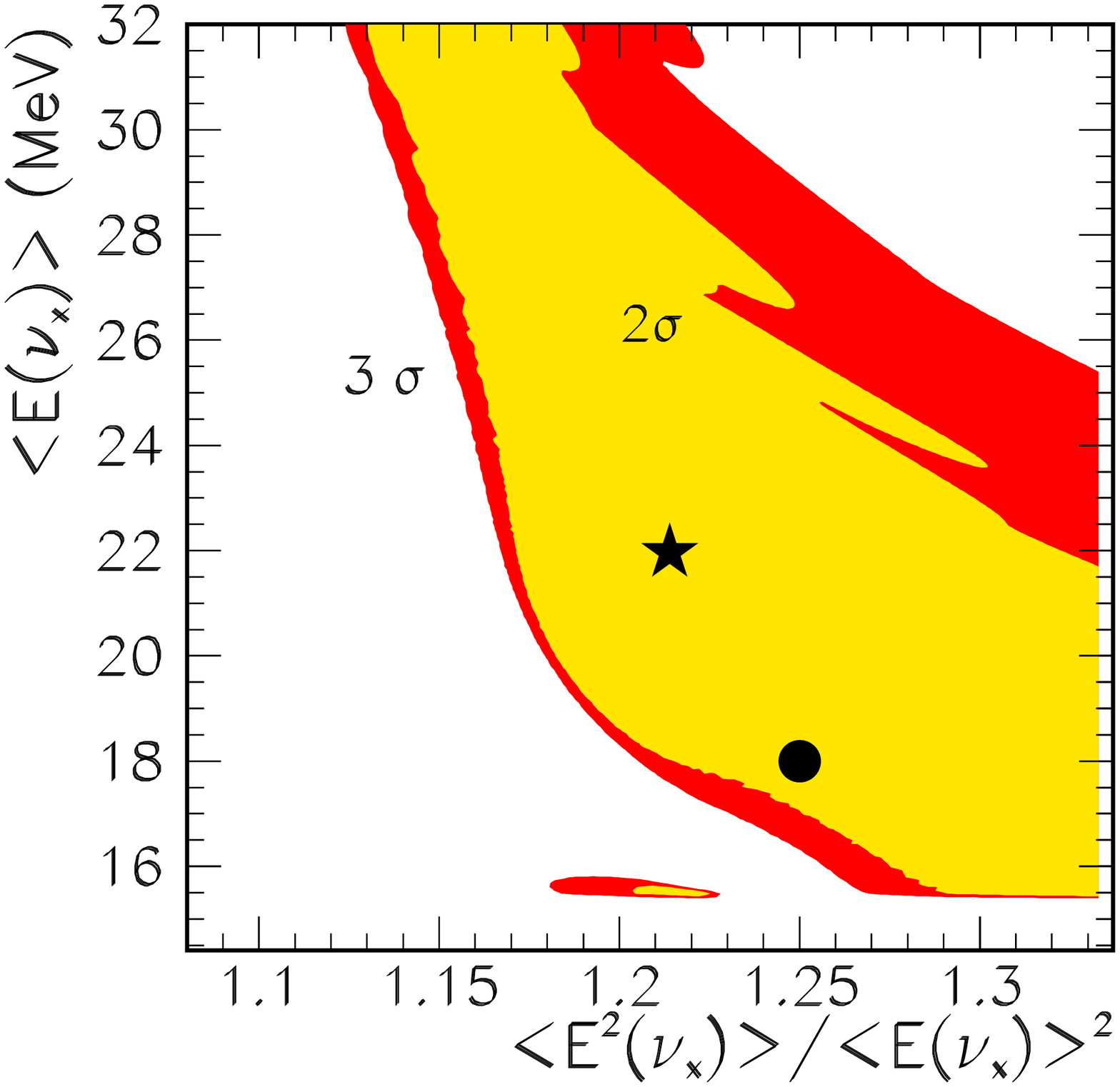}
    \includegraphics[width=0.33\textwidth,angle=0]{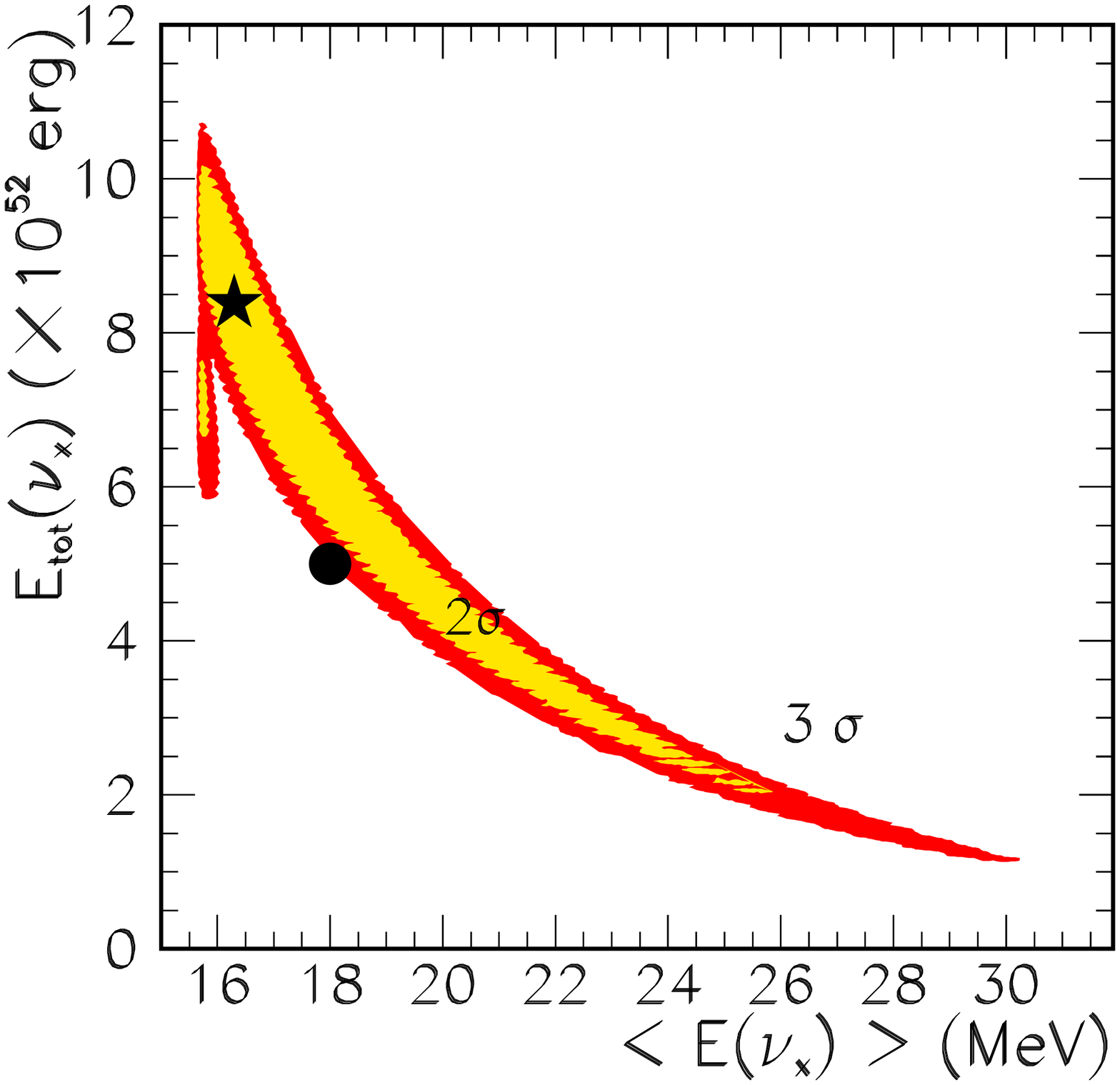}
    \includegraphics[width=0.33\textwidth,angle=0]{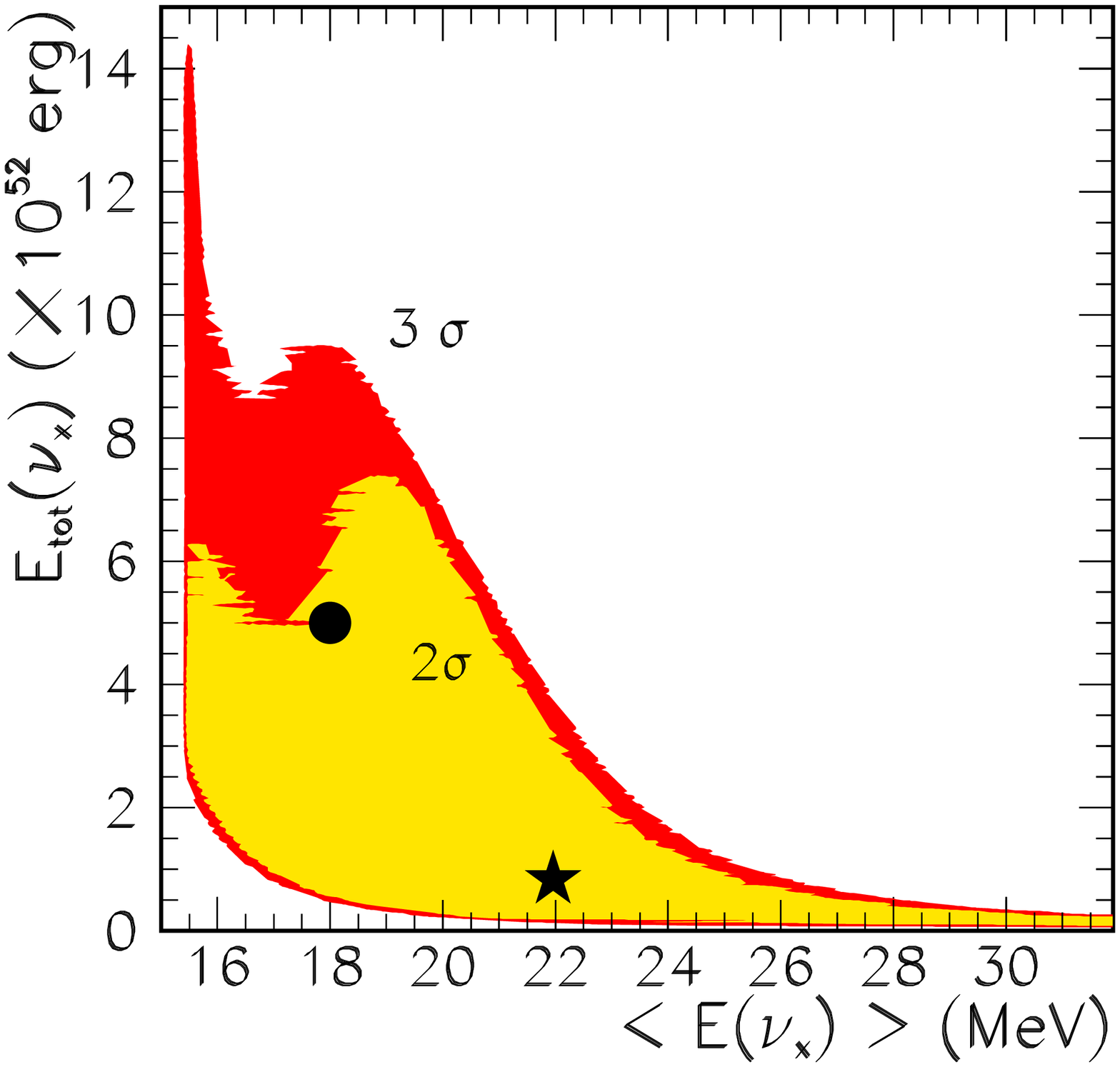}
  \end{center}
  \vglue -0.5cm
  \caption{Left panels: Allowed $2\sigma$ CL and $3\sigma$ CL contours
    assuming data generated with a Fermi-Dirac distribution and fitted
    with the Garching parametrization. Right panels: Allowed
    $2\sigma$ CL and $3\sigma$ CL contours assuming data generated
    with the Garching distribution and fitted using the Fermi-Dirac
    parametrization.  The initial values are taken from
    Tab.~\ref{tab:input} and are represented by a filled circle. The
    best fit points are shown with a star and have a value of $\chi^2_{\rm
      best}=5.29$ and $3.34$ in the left and right panels, respectively.  }
\label{fig:FD-Keil}
\end{figure}

The results of such a fit are presented in Fig.~\ref{fig:FD-Keil}. In
the left panels we show the allowed $2\sigma$ and $3\sigma$ CL
contours assuming data generated with a Fermi-Dirac distribution, with
input values given in Tab.~\ref{tab:input}, and fitted with the 
Garching parametrization. 
First of all, one realizes that the shape of the contours are very
similar to those presented in the previous sections, see
e.g. Figs.~\ref{fig:degeneracy1} and~\ref{fig:Keil-Keil}. In
particular, one finds the same one-dimensional 
 correlations between the
different parameters characterizing $F_{\nu_x}^0$. It is curious to
note that the size of the allowed regions is smaller than in the
previous figures.
The second feature to point out is that the best-fit point does not
coincide with the initial values assumed to generate the data. The
$\chi^2$ value obtained in the best-fit point is $\chi^2_{\rm bets}=
5.29$\footnote{ Notice, however, that these numbers have mild
  dependence on how fine is the mesh by which the parameter space is
  covered.  }.

In the right panels we show the same contours fixing the data with the
Garching parametrization and carrying out the fit assuming a
Fermi-Dirac distribution.
One can still recognize roughly the same tendencies as in the previous
case. However the contours are much bigger. As far as the best fit
point is concerned, it has a value of $\chi^2_{\rm bets}= 3.34$, and
as in the previous case it also does not coincide with the input
values.  These broad features indicate that Fermi-Dirac
parametrization has somewhat higher {\em flexibility} than the
Garching one, at least for the input values we used in this paper.

We note here that the results presented in Fig.~\ref{fig:FD-Keil}
not only demonstrate the robustness of the
degeneracy but also serve for providing some clues to answering the
question of whether the SN neutrino observation can discriminate
between different energy distribution of SN flux.
The values of $\chi^2_{\text{best}}$ for both cases (input given by
Garching parametrization fitted by Fermi Dirac one, and vice versa)
are small compared to the number data points (bins), 85 bins.  This
implies that both distributions can fit equally well irrespective of
the assumed input. Within our current procedure it seems rather
difficult, by fitting only the observed data, to say which
parametrization would describe better the SN spectra.
We note that such a difficulty arises because the observed 
spectra are given by the superposition of the 2 spectra which 
is the origin of the degeneracy we found in this work. 
If the spectra were described only by the single spectrum 
by either Fermi-Dirac or Garching parametrization 
(hence no degeneracy) it would be possible to distinguish 
the functional form thanks to the huge statistics. 

\vglue 1cm

\subsection{ Analytic understanding of correlations } 
\label{subsec:understanding}

Examining previous figures, e.~g. Fig.~\ref{fig:degeneracy1},
Fig.~\ref{fig:Keil-Keil} and Fig.~\ref{fig:FD-FD}, one sees that they
exhibit some intriguing correlations between the flux parameters.  Let
us now try to understand these features.
We have seen in Fig.~\ref{fig:example} that the degeneracy is present
at the level of the fluxes arriving at the Earth, already before the
detection process enters into the game.  In particular, since our
detection channel is the inverse beta decay, all the features observed
in the fit should be encoded in the $\bar\nu_e$ flux at the Earth.
For this reason let us consider the following quantities 
\begin{eqnarray}
\label{eq:moments}
E^{\rm tot}_{\rm ob} = \int E F_{\bar\nu_e}{\rm d}E,  
\hspace{1cm}
\langle E_{\rm ob} \rangle = \frac{\int E F_{\bar\nu_e}{\rm d}E}
{\int F_{\bar\nu_e}{\rm d}E}, 
\hspace{1cm} 
\langle E^2_{\rm ob} \rangle = 
\frac{\int E^2 F_{\bar\nu_e}{\rm d}E}{\int F_{\bar\nu_e} {\rm d}E} 
\end{eqnarray}
These functionals are determined in terms of the energy spectrum of
$F_{\bar\nu_e}$ at the Earth.

Of the six fit parameters we have, one notices from the previous
Figures that the range of variation of two of them is rather narrow,
namely $\langle E_{\bar\nu_e} \rangle $ and $p_{\bar\nu_e}$. Hence in
order to understand the nature of the correlations we need only the 4
remaining ``effective'' parameters.
Using Eq.~(\ref{eq:moments}) and requiring these quantities to be
equal to the observed (input) values, we see that these 4 parameters
are correlated, and one can choose to express the dependence of
$E^{\rm tot}_{\bar\nu_e},~\langle E_{\nu_x}\rangle$, and $p_{\nu_x}$,
in terms of one, e.g. $E^{\rm tot}_{\nu_x}$.

Taking into account the relation in Eq.~(\ref{eq:F_A}),
$E_{\rm ob}^{\rm tot}$ is trivially obtained by our assumption on the
initial values of the total energies $E^{\rm tot}_{\bar\nu_e} = E^{\rm
  tot}_{\nu_x} = 5\times 10^{52}~{\rm erg}$, see Tab.~\ref{tab:input},
as
\begin{equation}
E^{\rm tot}_{\rm ob} = c^2_{12} E^{\rm tot}_{\bar\nu_e} + s^2_{12}
E^{\rm tot}_{\nu_x} =
5\times 10^{52}~{\rm erg}\,,
\label{eq:Etot_ob}
\end{equation}
where $c^2_{12}$ and $s^2_{12}$ stand for $\cos^2\theta_{12}$ and
$\sin^2\theta_{12}$, respectively.
If we now fix $E^{\rm tot}_{\rm ob}$ to this value and let $ E^{\rm
  tot}_{\bar\nu_e}$ and $E^{\rm tot}_{\nu_x}$ freely vary we get,
after substituting the value of $\theta_{12}$ the
following trivial relation
\begin{equation}
E^{\rm tot}_{\bar\nu_e} = 1.43 E^{\rm tot}_{\rm ob}-0.43 E^{\rm
  tot}_{\nu_x}\, ,
\label{eq:Etot_ob2}
\end{equation}
which explains the anti-correlation found between $E^{\rm
  tot}_{\bar\nu_e}$ and $E^{\rm tot}_{\nu_x}$ in the right panel
in the second row of Fig.~\ref{fig:Keil-Keil}.

\begin{figure}[ht!]
  \begin{center}
    \includegraphics[width=0.55\textwidth,height=.45\textheight]{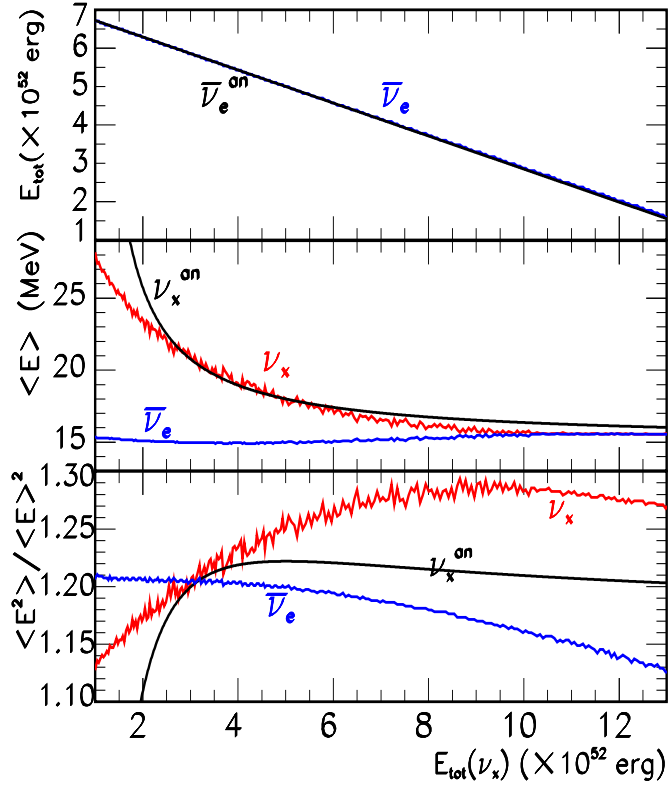}
  \end{center}
\vglue -0.8cm
\caption{Various quantities are plotted as a function of $E^{\rm
    tot}_{\nu_x}$, in blue and red for $\bar\nu_e$ and $\nu_x$,
  respectively. The total energy $E^{\rm tot}_{\bar\nu_e}$ is shown in
  the top panel, the average energies $\langle E\rangle$ in the middle
  panel, and the width parameters $\langle E^2\rangle / \langle
  E\rangle ^2$ in the bottom one. The analytical values predicted in
  Eqs.~(\ref{eq:Etot_ob2},~\ref{eq:Eav_ob2},~\ref{eq:E2av_ob2}) are
  also plotted in black. See text for details.  }
  \label{fig:1dprojection}
\end{figure}

The approximate quasi one-dimensional correlation is displayed in
Fig.~\ref{fig:1dprojection}.  In the figure, we plot 
$\alpha=\alpha(E^{\rm tot}_{\nu_x})$, where $\alpha$ represents
$E^{\rm tot}_{\bar\nu_e}$, $\langle E_{\bar\nu_e} \rangle$, $\langle
E_{\nu_x} \rangle$, $p_{\bar\nu_e}$, and $p_{\nu_x}$.  It is defined
as the value of $\alpha$ that minimizes $\chi^2(\alpha,E^{\rm
  tot}_{\nu_x})$ for a given $E^{\rm tot}_{\nu_x}$.  This plot
illustrates how the different parameters adjust themselves when
$E^{\rm tot}_{\nu_x}$ is varied so that the new set of parameters fits
equally well. Together with $\alpha(E^{\rm tot}_{\nu_x})$ we show the
analytical expression obtained by requiring the quantities defined in
Eq.~(\ref{eq:moments}) to be equal to the observed (input) values.  

In the top panel of Fig.~\ref{fig:1dprojection} we can see how the
expected anti-correlation for $\alpha=E^{\rm tot}_{\bar\nu_e}$ (see
the right panel in the second row in Fig.~\ref{fig:Keil-Keil})
completely agrees with the analytical expression~\ref{eq:Etot_ob2}.

Let us now consider the mean energy of the $\bar\nu_e$ arriving at the
Earth. According to the values assumed for the input parameters
describing the flux one expects, 
\begin{equation}
\langle E_{\rm ob} \rangle = \frac{E^{\rm tot}_{\rm ob}}{c^2_{12}
  E^{\rm tot}_{\bar\nu_e}/\langle E_{\bar\nu_e}\rangle  + s^2_{12}
  E^{\rm tot}_{\nu_x}/\langle E_{\nu_x}\rangle} = 15.79~{\rm MeV}\,.
\label{eq:Eav_ob}
\end{equation}
We can now proceed as before and rewrite $\langle E_{\nu_x} \rangle$
in terms of $E^{\rm tot}_{\nu_x}$ taking into account
Eq.~(\ref{eq:Etot_ob}):
\begin{equation}
\langle E_{\nu_x} \rangle = \frac{s^2_{12} E^{\rm
    tot}_{\nu_x}}{s^2_{12} E^{\rm tot}_{\nu_x} - E^{\rm tot}_{\rm
    ob}(1-\langle E_{\bar\nu_e} \rangle/\langle E_{\rm ob}
    \rangle)}\langle E_{\bar\nu_e} \rangle = 15\frac{E^{\rm
    tot}_{\nu_x}}{E^{\rm tot}_{\nu_x} - 8.34\times 10^{51}~{\rm
    erg}}~{\rm MeV} \,.
\label{eq:Eav_ob2}
\end{equation}
where in the last step we have assumed that $\langle E_{\bar\nu_e}
\rangle$ and $\langle E_{\rm ob} \rangle$ are constant and equal to 15
MeV and 15.79 MeV, respectively. 
Under this assumption we expect that $\langle E_{\nu_x}\rangle$
decreases for large values of $E^{\rm tot}_{\nu_x}$.  This behavior
can be clearly seen in the right panel in Fig.~\ref{fig:degeneracy1}
as well as in the middle panel of Fig.~\ref{fig:1dprojection} (see the
red solid curve).  Moreover, from the same panel one sees that the
assumption of constant $\langle E_{\bar\nu_e}\rangle$ (blue curve) is
justified when compared to the variation of $\langle E_{\nu_x}\rangle$
(red curve). However, this assumption is not perfect and therefore the
observed variation of $\langle E_{\nu_x}\rangle$ and the analytical
prediction from Eq.~\ref{eq:Eav_ob2} (black curve) do not coincide.

Now we analyze the second moment of the flux, 
\begin{equation}
\langle E^2_{\rm ob} \rangle = \frac{c^2_{12} E^{\rm
  tot}_{\bar\nu_e} \langle E_{\bar\nu_e}\rangle p_{\bar\nu_e} + s^2_{12} E^{\rm
  tot}_{\nu_x} \langle E_{\nu_x}\rangle p_{\nu_x}}{c^2_{12}
  E^{\rm tot}_{\bar\nu_e}/\langle E_{\bar\nu_e}\rangle  + s^2_{12}
  E^{\rm tot}_{\nu_x}/\langle E_{\nu_x}\rangle} = 305.53~{\rm MeV}^2\,.
\label{eq:E2av_ob}
\end{equation}
Taking into account Eqs.~(\ref{eq:Etot_ob}), (\ref{eq:Eav_ob})
and~(\ref{eq:E2av_ob}) we can express $p_{\nu_x}$ in terms of
$\langle E_{\nu_x} \rangle$:
\begin{equation}
p_{\nu_x} = \frac{\langle E^2_{\rm ob} \rangle}{\langle E_{\bar\nu_e}
  \rangle \langle E_{\nu_x} \rangle}
\left\{ \left(1-\frac{\langle E_{\bar\nu_e}\rangle}{\langle E_{\nu_x}
  \rangle}\right) \frac{\langle E_{\bar\nu_e}\rangle/\langle E_{\rm
  ob} \rangle - \langle E^2_{\bar\nu_e}\rangle/\langle E^2_{\rm ob}
  \rangle}{1-\langle E_{\bar\nu_e}\rangle/\langle E_{\rm ob}
  \rangle} + \frac{\langle E^2_{\bar\nu_e} \rangle}{ \langle E^2_{\rm ob} \rangle}
\right\}\,.
\label{eq:E2av_ob2}
\end{equation}
Analogously to the previous case we can require $\langle E^2_{\rm ob}
\rangle$ and $p_{\bar\nu_e}$ to be constant.  Then, from this
expression one can see how as $\langle E_{\nu_x} \rangle$ increases
the $\nu_x$ spectrum has to become more pinched in order to keep the
same {\em shape}. Taking into account that $\langle E_{\nu_x}\rangle$
and $E^{\rm tot}_{\nu_x}$ are anti-correlated one expects a {\it
  positive} correlation between $p_{\nu_x}$ and $E^{\rm
  tot}_{\nu_x}$. This is exactly what is seen in
Figs.~\ref{fig:Keil-Keil} (left panel in the second row) and in the bottom panel
of Fig.~\ref{fig:1dprojection} (red curve).  In the latter figure we
see how this argument is mainly qualitative, since $p_{\bar\nu_e}$
is not constant over the whole range of the $E^{\rm tot}_{\nu_x}$
considered. Actually, when comparing this result with the analytical
condition (\ref{eq:E2av_ob2}) one realizes that this qualitative
argument only applies for a definite range of $E^{\rm tot}_{\nu_x}$.

\section{Resolving the degeneracy?}
\label{sec:solveDG}

Once the presence and robustness of the degeneracy have been
established, the next step is to point out strategies to solve the
problem. Here we comment on three possible directions that could be
followed.

First from the theory side we note that a better understanding of the
neutrino flux formation would narrow down the range of values that the
flux parameters could take. In particular we have seen in
Fig.~\ref{fig:Keil-Keil} how our knowledge of the pinching parameter
can play an essential role in reducing the uncertainties on the
traditionally considered parameters $\langle E_{\nu_\alpha}\rangle$
and $E^{\rm tot}_{\nu_\alpha}$.

From an experimental point of view one may correctly argue that the
degeneracy we have pointed out is an artifact of our treatment, which
uses only the $\bar{\nu}_{e}$ absorption reaction.  The obvious way of
resolving such a degeneracy would be to include other reactions into
the analysis, such as neutrino elastic scattering off electrons, or CC
and NC reactions with oxygen.
Indeed, the quasi one-dimensional nature of the degeneracy we found
would suggest that it would be lifted by adding an extra high
statistics observable, independent of inverse beta decay.

However, the situation in our case is quite different.  To understand
why, one must recall the following crucial fact. In order to include
the previous reactions one must also include $\nu_{e}$ into the
analysis.  As a result instead of the six supernova neutrino flux
parameters used in our fitting procedure we would need nine.  Since
the number of events due to the above reactions is more than one order
of magnitude less than those coming from the $\bar{\nu}_{e}$
absorption reaction, it is not obvious at all that, by adding these
other reactions one will indeed resolve the degeneracy we have
uncovered.

If the degeneracy cannot be resolved simply by using water Cherenkov
detectors one must think of possible alternative detectors to combine
with.  
For example the situation might improve either by: (1) adding
  detectors with better sensitivity on $\nu_{x}$ parameters, or (2)
  adding detectors sensitive to $\nu_{e}$.
Good candidates in the current and near future experiments 
which can contribute towards these goals would be 
Borexino~\cite{Cadonati:2000kq}, KamLAND (in its future 
solar neutrino observation phase)~\cite{Suekane:2006qi}
and HALO project~\cite{HALO}. 
The former 2 detectors will be able to observe proton recoil 
in $\nu p$ elastic  scattering in a liquid scintillator 
~\cite{Beacom:2002hs}.
We stress that this would provide a unique
opportunity to obtain the spectrum information of NC reactions which
should be very important in resolving the degeneracy.
The latter, on the other hand, would have capabilities 
to detect $\nu_{x}$ and $\nu_{e}$  through the NC and 
the CC reactions on lead.  
With regard to the possibility (2) above it would be interesting to
consider a high statistics measurement of $\nu_e$ events in a liquid
argon neutrino detector through the charged current absorption of
$\nu_e$ by $^{40}$Ar \cite{GilBotella:2004bv}.
As can be inferred from
Eq.~(\ref{eq:F_A}), the dependence of the $\nu_e$
flux at the Earth on the original $F^0_{\nu_e}$ and $F^0_{\nu_x}$ is
different from that of antineutrinos. In particular for {\em large}
$\theta_{13}$ the following condition holds
\begin{equation}
F_{\nu_e} =
s^2_{13} F_{\nu_e}^0 + c^2_{13} F_{\nu_x}^0 
\approx F_{\nu_x}^0 \,.
\end{equation}
Therefore, one could hope that this fact could break the degeneracy
expected in water Cherenkov detectors.  Nevertheless the possibility
of distinguishing between the $\nu_e$ and $\bar\nu_e$ signals is in
any case non-trivial.
Therefore, a realistic nine-dimensional parameter analysis taking into
account all neutrino flavors is outside the scope of this paper but
should be taken up as a next step.

\section{Concluding remarks}
\label{conclusion}

In this paper, we have reexamined the possibility of reconstructing
the initial fluxes of the neutrinos emitted in a future galactic
core-collapse supernova by using a Megaton-sized water Cherenkov
detector.

The three parameters that are usually considered to characterize the
non-thermal supernova neutrino flux are the average and total energies
of each species, and the so-called pinching parameter. The latter is
connected to the second moment 
of the distribution function and modulates its shape.
Due to the current uncertainty on its precise value, we have included
the pinching parameter as a fit parameter, in contrast to previous
works where only the average and total energies were considered.

In order to illustrate our results we have considered the following
scenario. First of all, from all reactions giving rise to the neutrino
signal in a water Cherenkov detector we have concentrated on the
inverse beta decay, $\bar\nu_e + p \to e^+ + n$, which is by far the
most important one. 
Therefore, our setting is sensitive only to the anti-neutrino fluxes
at terrestrial detectors.

As far as the neutrino properties are concerned, we have focused on
the case of normal mass hierarchy neutrinos, neglecting also any
non-standard neutrino interactions which could affect neutrino
propagation in an important way. Such well-defined scenario is much
less affected by uncertainties due to neutrino self-interaction in the
inner layers or the distortion of the matter density due to the shock
wave passage.

On the other hand the $\bar\nu_e$ flux arriving at the Earth is in
this case a strong mixture of the initial $\bar\nu_e$ and
$\nu_x$. Therefore the detected $\bar\nu_e$ flux would provide us with
information not only about $\bar\nu_e$ but also $\nu_x$.
Under these assumptions we performed a $\chi^2$ analysis on
artificially generated data.

We found that the inclusion of the pinching parameter in the fit
analysis has a drastic consequence: the appearance of a continuous
degeneracy in the determination of the $\nu_x$ flux parameters. This
degeneracy is quite robust, as it persists irrespective of the
parametrization taken for the neutrino distributions.
This makes very difficult a complete determination of the different
parameters characterizing the $\nu_x$ flux using the inverse beta
decay events, even in the case of a Megaton water Cherenkov detector.

The solution to this degeneracy must come from a better understanding
of the neutrino spectra formation, as well as an optimization of the
information provided by the complementary neutrino reactions involved
not only in water Cherenkov detectors, but also in alternative
detector techniques.

Note added:
After completion of our work we became aware of the paper
\cite{Skadhauge:2008gf}, which was in fact triggered by previous
private communications with us. Their analysis includes other
reactions available in water Cherenkov detectors.
However, their results indicate that the addition of these channels
is not enough to fully resolve the degeneracy pointed out here.

\section*{Acknowledgments}

One of the authors (H.M.) thanks Carlos Pe\~na-Garay and 
Jos\'e Furtado Valle for their generous supports which made 
his multiple visits to IFIC, University of Valencia, possible. 
One of the authors (R.T.) is grateful to Japan Society for the Promotion 
of Science for support to his visit to Tokyo Metropolitan University in 2006 
by providing a Postdoctoral Fellowships for Foreign Researchers.
RT was also supported by the Juan de la Cierva programme, an ERG from the
European Commission.
This work was supported by Spanish grants FPA2005-01269,
FPA2005-25348-E (MEC), and FPA2008-00319/FPA, ACOMP07/270 (Generalitat
Valenciana), European Commission Contracts RII3-CT-2004-506222
(ILIAS/N6), MRTN-CT-2004-503369, in part by KAKENHI, Grant-in-Aid for
Scientific Research (B), Nos. 16340078 and 19340062, Japan Society for
the Promotion of Science, Funda\c{c}\~ao de Amparo \`a Pesquisa do
Estado de Rio de Janeiro (FAPERJ) and Conselho Nacional de Ci\^encia e
Tecnologia (CNPq).


\end{document}